\documentclass[12pt]{article}

\usepackage[dvips]{graphicx}

\newcommand{\tsc}[1]{\textsc{#1}}
\newcommand{\tbf}[1]{\textbf{#1}}
\newcommand{\ttt}[1]{\texttt{#1}}

\newcommand{\Py}{\tsc{Pythia}}
\newcommand{\Mo}{\tsc{Moncher}}

\newcommand{\reg}{\mathrm{I}\!\mathrm{R}}

{\end{list}}
\newcounter{enumct}

\newenvironment{entry}%
{\begin{list}{}{\setlength{\topsep}{0mm} \setlength{\itemsep}{0mm}
\setlength{\parskip}{0mm} \setlength{\parsep}{0mm}
\setlength{\leftmargin}{20mm} \setlength{\rightmargin}{0mm}
\setlength{\labelwidth}{18mm} \setlength{\labelsep}{2mm}}}%
{\end{list}}
\newenvironment{subentry}%
{\begin{list}{}{\setlength{\topsep}{0mm} \setlength{\itemsep}{0mm}
\setlength{\parskip}{0mm} \setlength{\parsep}{0mm}
\setlength{\leftmargin}{10mm} \setlength{\rightmargin}{0mm}
\setlength{\labelwidth}{18mm} \setlength{\labelsep}{2mm}}}%
{\end{list}}
\newcommand{\itemc}[1]{\item[\textbf{#1}\hfill]}
\newcommand{\iteme}[1]{\item[\texttt{#1}\hfill]}

\newcommand{\drawbox}[1]{\vspace{\baselineskip}\noindent%
\fbox{\texttt{#1}}\vspace{0.5\baselineskip}}
\newcommand{\drawboxtwo}[2]{\vspace{\baselineskip}\noindent%
\fbox{\begin{minipage}{150mm}\begin{tabbing}%
{\texttt{#1}}\\{\texttt{#2}}%
\end{tabbing}\end{minipage}}\vspace{0.5\baselineskip}}
\newcommand{\drawboxthree}[3]{\vspace{\baselineskip}\noindent%
\fbox{\begin{minipage}{150mm}\begin{tabbing}%
{\texttt{#1}}\\{\texttt{#2}}\\{\texttt{#3}}%
\end{tabbing}\end{minipage}}\vspace{0.5\baselineskip}}
\newcommand{\drawboxfour}[4]{\vspace{\baselineskip}\noindent%
\fbox{\begin{minipage}{150mm}\begin{tabbing}%
{\texttt{#1}}\\{\texttt{#2}}\\{\texttt{#3}}\\{\texttt{#4}}%
\end{tabbing}\end{minipage}}\vspace{0.5\baselineskip}}
\newcommand{\drawboxfive}[5]{\vspace{\baselineskip}\noindent%
\fbox{\begin{minipage}{150mm}\begin{tabbing}%
{\texttt{#1}}\\{\texttt{#2}}\\{\texttt{#3}}\\{\texttt{#4}}\\%
{\texttt{#5}}%
\end{tabbing}\end{minipage}}\vspace{0.5\baselineskip}}
\newcommand{\drawboxsix}[6]{\vspace{\baselineskip}\noindent%
\fbox{\begin{minipage}{150mm}\begin{tabbing}%
{\texttt{#1}}\\{\texttt{#2}}\\{\texttt{#3}}\\{\texttt{#4}}\\%
{\texttt{#5}}\\{\texttt{#6}}%
\end{tabbing}\end{minipage}}\vspace{0.5\baselineskip}}
\newcommand{\drawboxseven}[7]{\vspace{\baselineskip}\noindent%
\fbox{\begin{minipage}{150mm}\begin{tabbing}%
{\texttt{#1}}\\{\texttt{#2}}\\{\texttt{#3}}\\{\texttt{#4}}\\%
{\texttt{#5}}\\{\texttt{#6}}\\{\texttt{#7}}%
\end{tabbing}\end{minipage}}\vspace{0.5\baselineskip}}
\newcommand{\drawboxnine}[9]{\vspace{\baselineskip}\noindent%
\fbox{\begin{minipage}{150mm}\begin{tabbing}%
{\texttt{#1}}\\{\texttt{#2}}\\{\texttt{#3}}\\{\texttt{#4}}\\%
{\texttt{#5}}\\{\texttt{#6}}\\{\texttt{#7}}\\{\texttt{#8}}\\%
{\texttt{#9}}%
\end{tabbing}\end{minipage}}\vspace{0.5\baselineskip}}
\newcommand{\boxsep}{\vspace{0.5\baselineskip}} 
 
\newlength{\captivewidth}
\newlength{\tablinsep}
\newlength{\halfpagewid}
\newlength{\abstwidth}
\begin{document}
\sloppy
 
\pagestyle{empty}
 
\begin{flushright}
CERN\\
June 2011
\end{flushright} 
\vskip 1.75cm

\begin{center}

\tbf{{\Large\bf M}{\large\bf ONCHER:}}\\[3mm]
\tbf{{\Large\bf M}{\large\bf ONte Carlo generator for CHarge Exchange Reactions}}\\[3mm]
\tbf{{\large\bf Version 1.1}}\\[10mm]
\tbf{{\large\bf Physics and Manual}} \\[20mm]
{ R.A.~Ryutin, A.E.~Sobol, V.A.~Petrov  } \\[7mm]
{\small Institute for High Energy Physics} \\
{\small {\it 142 281} Protvino, Russia}

\vskip 1.75cm
{\bf
\mbox{Abstract}}
  \vskip 0.3cm

\newlength{\qqq}
\settowidth{\qqq}{In the framework of the operator product  expansion, the quark mass dependence of}
\hfill
\noindent
\begin{minipage}{\qqq}

\Mo\ is a Monte Carlo event generator for simulation of single and double charge exchange reactions 
in proton-proton collisions at energies from 0.9 to 14 TeV. Such reactions, $pp\to n+X$ and 
$pp\to n+X+n$, are characterized by leading neutron production. They are dominated by $\pi^+$ exchange
and could provide us with more information about total and elastic $\pi^+ p$ and 
$\pi^+\pi^+$ cross sections and parton distributions in pions in the still unexplored 
kinematical region.  

\end{minipage}
\end{center}


\begin{center}
\vskip 0.5cm
{\bf
\mbox{Keywords}}
\vskip 0.3cm

\settowidth{\qqq}{In the framework of the operator product  expansion, the quark mass dependence of}
\hfill
\noindent
\begin{minipage}{\qqq}
Single Charge Exchange -- Double Charge Exchange -- pion-proton -- 
pion-pion -- cross sections -- event generator
\end{minipage}

\end{center}

\clearpage

\tableofcontents
 
\clearpage

\pagestyle{plain}
\setcounter{page}{1}

\section{Introduction}

In the paper we present a new  Monte-Carlo event generator \Mo. The generator is devoted
to the simulation of single and double charge exchange reactions in proton-proton 
collisions at energies from 0.9 to 14 TeV. This region of energies covers the present 
capabilities of  the LHC. Charge exchange reactions, $pp\to n+X$ and $pp\to n+X+n$, are 
characterized by the leading neutron production. They can be studied with LHC detectors 
incorporated with forward neutron calorimeters like the ZDC (Zero Degree Calorimeter)~\cite{ZDC} 
in the CMS~\cite{CMS}. 

Reactions with the leading neutron production are dominated by $\pi^+$ 
exchange~\cite{KMRn1}-\cite{workn2}. At the LHC they could provide us with information about 
$\pi^+ p$ and $\pi^+ \pi^+$ interactions in the  region of energies 1-5 TeV in the c.m.s. 
Using indirect methods~\cite{ourneutrontot}-\cite{ourneutronel} we could extract total and elastic 
$\pi^+ p$ and $\pi^+ \pi^+$ cross sections at these energies. It is worth mentioning that  
the total cross-section of $\pi^+ p$ interaction is measured only at energies up to 25 GeV by direct methods in 
the fixed target experiments~\cite{PDG} and total and elastic cross sections of $\pi^+ \pi^+$ 
interactions are extracted from the data at energies 1.5-18.4 GeV only 
(see Ref.~\cite{pipiextract}-\cite{pipiextract3}). 
Moreover, a study of charge exchange reactions with hard scattering $\pi^+ p$ and $\pi^+ \pi^+$ 
followed by dijet production at the LHC, could  provide us with parton distributions in the pion 
in the unexplored kinematical domain. 
So, we had weighty motivations to develop a model and to create a generator for charge exchange
simulation which could be used  at high energies of the LHC.

An important point is that at high energies we have to take into account effects of soft rescattering which can be 
calculated  as corrections to the Born approximation. In the calculations of such absorptive 
effects we use the Regge-eikonal approach~\cite{3Pomerons}. For  $\pi^+ p$ and $\pi^+ \pi^+$ 
interactions several models which predict different cross sections have been applied. In addition 
to the dominant $\pi^+$ exchange  we have calculated contributions of two other important Reggeons, 
$\rho^+$ and $a_2^+$, to the charge exchange cross section~\cite{ourneutronregg} and implemented 
both Reggeons to the generation. \Py\ 6.4\ \cite{pythia} is used as a basic generator for \Mo. 
\Mo\ has the same format of events, parameters and common blocks as \Py\ . \Py subroutines 
are used also for the simulation of $\pi^+ p$ and $\pi^+ \pi^+$  interactions and for the subsequent 
hadronization and decays.


\section{Physics Overview}
\label{s:physics}

\subsection{Single Pion Exchange}
\label{ss:SPE}

\begin{figure}[t!]
\begin{center} 
\includegraphics[width=0.7\textwidth]{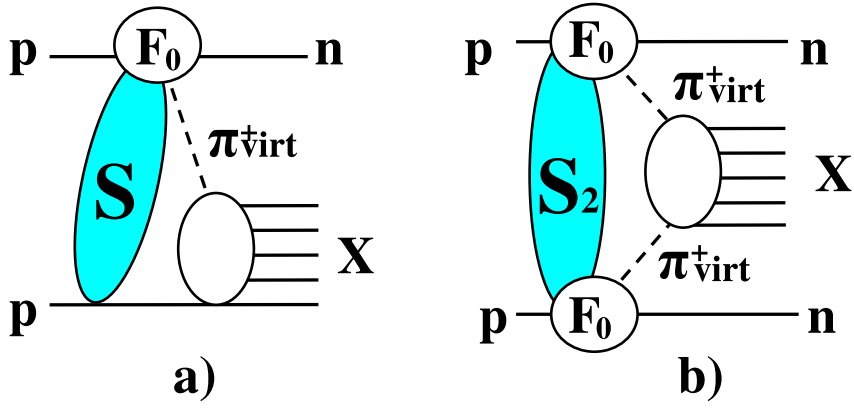}
\caption{\label{fig:1} 
Amplitudes of the processes: a) $p+p\to n+X$ (S$\pi$E), 
b) $p+p\to n+X+n$ (D$\pi$E). $S$ and $S_2$ represent soft rescattering corrections. 
}
\end{center} 
\end{figure}

The diagram of the Single pion Exchange (S$\pi$E) process $p+p\to n+X$ is presented in 
Fig.~\ref{fig:1}a. The momenta are $p_1$, $p_2$, $p_n$, $p_X$ respectively. 
In the center-of-mass frame these can be represented as follows 
(boldface letters denote transverse momenta):
 \begin{equation}
 \label{kin1a}p_1=\left(\frac{\sqrt{s}}{2},\frac{\sqrt{s}}{2}\beta ,\mbox{\bf 0}\right),\; 
 p_2=\left(\frac{\sqrt{s}}{2},-\frac{\sqrt{s}}{2}\beta ,\mbox{\bf 0}\right).
\end{equation}
With this notation, the momentum of the $\pi$ is
\begin{equation}
\label{kin1c}p_{\pi}=\left( \xi\frac{\sqrt{s}}{2}\beta^2+\frac{t+m_p^2-m_n^2}{\sqrt{s}},
\xi\frac{\sqrt{s}}{2}\beta ,\mbox{\bf q}\right),\\
\end{equation}
and
\begin{eqnarray}
&&\label{kin1d}p_n=p_1-p_{\pi},\\
&&p_X^2=M^2,
\end{eqnarray}
\begin{eqnarray}
&&\label{kin1e}\xi=\frac{M^2-m_n^2-2(t+m_p^2-m_n^2)}{s\beta^2}\simeq\frac{M^2}{s},\\
&&\label{kin1f}-t=\frac{\mbox{\bf q}^2+\xi^2\beta^2m_p^2+(m_n^2-m_p^2)
\left( \xi\beta^2-\frac{m_n^2-m_p^2}{s}\right)}{1-\xi\beta^2+\frac{2(m_n^2-m_p^2)}{s}}
\simeq\frac{\mbox{\bf q}^2+\xi^2m_p^2}{1-\xi},\\
&& \label{kin1g}\beta=\sqrt{1-\frac{4m_p^2}{s}}.
 \end{eqnarray}

As a Born approximation for $\pi$ exchange we use the familiar triple-Regge formula. 
This formula can be rewritten as 
\begin{eqnarray}
&&\frac{d\sigma_{X,{\rm S}\pi {\rm E}}}{d\xi dt\; d\Phi_X}=\frac{G_{\pi^+pn}^2}{16\pi^2}
\frac{-t}{(t-m_{\pi}^2)^2} F_0^2(t) \xi^{1-2\alpha_{\pi}(t)} \nonumber\\
&&\label{born}\times\frac{d\sigma_{X,\pi^+ p}(\xi s)}{d\Phi_X} S(s/s_0, \xi, t),
\end{eqnarray}
where $\Phi_X$ is the phase space for the system X produced in the $\pi^+ p$ scattering, the pion trajectory 
is $\alpha_{\pi}(t)=\alpha^{\prime}_{\pi}(t-m_{\pi}^2)$. The slope $\alpha^{\prime}\simeq 0.9$~GeV$^{-2}$, 
$\xi=1-x_L$, were $x_L$ is the fraction of the initial proton longitudinal momentum carried by the neutron, 
and $G_{\pi^0pp}^2/(4\pi)=G_{\pi^+pn}^2/(8\pi)=13.75$~\cite{constG,constG2}. The form factor $F_0(t)$ is usually 
expressed as an exponential
\begin{equation}
\label{formfactor}
F_0(t)=\exp(bt),
\end{equation}
where, from recent data~\cite{HERA2},\cite{KMRn1c16}, we expect $b\simeq 0.3\; {\rm GeV}^{-2}$.
We are interested in the kinematical range 
 \begin{equation}
 \label{modlims}
 0.01\;{\rm GeV}^2<|t|<0.5\;{\rm GeV}^2,\;\xi<0.4,
 \end{equation}
where formula~(\ref{born}) dominates according to~\cite{KMRn1c13} and~\cite{KMRn1c14}. 
At high energies we can use any adequate parametrizations of different $\pi^+ p$ cross-sections. 

\subsubsection{Absorptive corrections}
\label{sss:absorption}
 
 The suppression factor $S$ arises from absorptive corrections~\cite{KMRn1}. 
 We estimate absorption in the initial state for inclusive reactions and for both initial 
 and final states in exclusive exchanges. For this task we use our model with 3 Pomeron 
 trajectories~\cite{3Pomerons}:
\begin{eqnarray}\label{3IPtrajectories}
&&\alpha_{IP_1}(t)-1= (0.0578\pm0.002)+(0.5596\pm0.0078)t \;,\nonumber\\
&&\alpha_{IP_2}(t)-1= (0.1669\pm0.0012)+(0.2733\pm 0.0056)t \;,\\
&&\alpha_{IP_3}(t)-1= (0.2032\pm0.0041)+(0.0937\pm0.0029)t \ .\nonumber
\end{eqnarray}
These trajectories  are the result of a 20 parameter fit of the total and differential cross-sections 
in the region 
$$
0.01\;{\rm GeV}^2<|t|<14\; {\rm GeV}^2,\; 8\; {\rm GeV}<\sqrt{s}<1800\; {\rm GeV}.
$$
Although the $\chi^2/d.o.f.=2.74$ is rather large, the model gives
good predictions for the elastic scattering (especially in the low-t region with $\chi^2/d.o.f.\sim1$). 

\begin{figure}[h!]
\begin{center} 
\includegraphics[width=\textwidth]{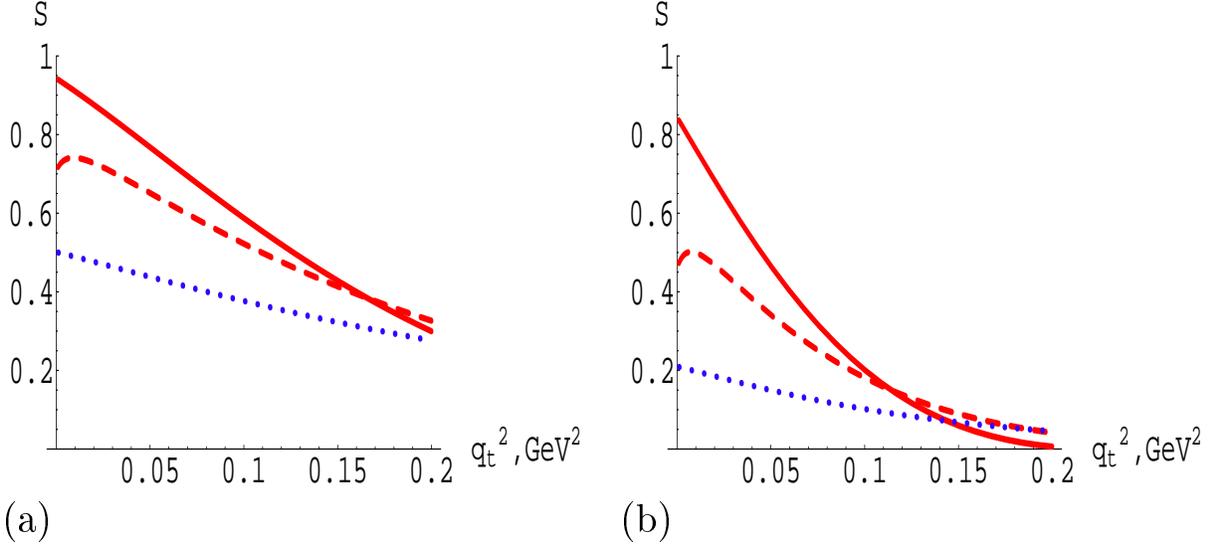}
\caption{\label{fig:2}Function $S(s/s_0,\xi,q_t)$ at 
a) $\sqrt{s}=62.7$~GeV and b) $\sqrt{s}=10$~TeV for different 
$\xi$ values: $\xi=0.3$ (dotted), $\xi=0.1$ (dashed) and $\xi=10^{-4}$ (solid).}
\end{center} 
\end{figure}

We use the procedure described in~\cite{workn1},\cite{workn2} to estimate the absorptive corrections. With an effective factorized form of (see hereunder) expression~(\ref{Udsigma}) used for convenience, we obtain:
\begin{eqnarray}
&& \label{Udsigma}\frac{d\sigma(s/s_0,\xi,\mbox{\bf q}^2)}{d\xi d\mbox{\bf q}^2}=S(s/s_0,\xi,\mbox{\bf q}^2) \frac{d\sigma_0(\xi,\mbox{\bf q}^2)}{d\xi d\mbox{\bf q}^2},\\
&& \frac{d\sigma_0(\xi,\mbox{\bf q}^2)}{d\xi d\mbox{\bf q}^2}=
\label{dsigma0qt}=(m_p^2\xi^2+\mbox{\bf q}^2)|\Phi_B(\xi,\mbox{\bf q}^2)|^2\frac{\xi}{(1-\xi)^2}\sigma_{\pi^+p}(\xi\;s),\\
&& \label{survival}S=\frac{m_p^2\xi^2 |\Phi_0(s/s_0,\xi,\mbox{\bf q}^2)|^2+\mbox{\bf q}^2|\Phi_s(s/s_0,\xi,\mbox{\bf q}^2)|^2}{(m_p^2\xi^2+\mbox{\bf q}^2)|\Phi_B(\xi,\mbox{\bf q}^2)|^2}.
\end{eqnarray}
The functions $\Phi_0$ and $\Phi_s$ arise from different spin contributions to the amplitude
\begin{equation}
\label{spinamplitude}
A_{p\to n}=\frac{1}{\sqrt{1-\xi}}\bar{\Psi}_n \left(
m_p\xi\; \hat{\sigma}_3\cdot \Phi_0+\mbox{\bf q}\;\hat{\vec{\sigma}}\cdot\Phi_s
\right) \Psi_p
\end{equation}
and both are equal to $\Phi_B$ in the Born approximation. Here $\hat{\sigma}_i$ are Pauli matrices and $\bar{\Psi}_n$, $\Psi_p$ are neutron and proton spinors. All the above functions can be calculated by the following set of formulae:
\begin{eqnarray}
&& \Phi_B(\xi,\mbox{\bf q}^2)=\frac{N(\xi)}{2\pi}\left( 
\frac{1}{\mbox{\bf q}^2+\epsilon^2}+\imath \frac{\pi\alpha_{\pi}^{\prime}}{2(1-\xi)}
\right)\exp(-\beta^2\mbox{\bf q}^2)\simeq\nonumber\\
&& \simeq \frac{N(\xi)}{2\pi}
\frac{1}{\mbox{\bf q}^2+\epsilon^2}
\frac{1}{1+\beta^2\mbox{\bf q}^2},\; \mbox{\bf q}\to 0,\\
&& N(\xi)=(1-\xi)\frac{G_{\pi^+pn}}{2}\xi^{\frac{\alpha_{\pi}^{\prime}\epsilon^2}{1-\xi}}\exp\left[ -b\frac{m_p^2\xi^2}{1-\xi}\right],\\
&&\beta^2=\frac{b+\alpha_{\pi}^{\prime}\ln\frac{1}{\xi}}{1-\xi},\; \epsilon^2=m_p^2\xi^2+m_{\pi}^2(1-\xi),\\
&& \Theta_0(b,\xi,|\mbox{\bf q}|)=\frac{b\; J_0(b|\mbox{\bf q}|)\left(K_0(\epsilon\;b)-K_0\left(\frac{b}{\beta}\right)\right)}{1-\beta^2\epsilon^2},\\
&& \Theta_s(b,\xi,|\mbox{\bf q}|)=\frac{b\; J_1(b|\mbox{\bf q}|)\left(\epsilon\; K_1(\epsilon\; b)-\frac{1}{\beta}K_1\left( \frac{b}{\beta}\right)\right)}{1-\beta^2\epsilon^2},\\
&& \Phi_0=\frac{N(\xi)}{2\pi}\int\limits_0^{\infty} db\; \Theta_0(b,\xi,|\mbox{\bf q}|)V(b),\\
&& |\mbox{\bf q}|\Phi_s=\frac{N(\xi)}{2\pi}\int\limits_0^{\infty} db\;\Theta_s(b,\xi,|\mbox{\bf q}|) V(b),
\end{eqnarray}
\begin{eqnarray}
&& V(b)=\exp\left( -\Omega_{el}(s/s_0,b)\right),\\
&& \label{U3}\Omega_{el}=\sum\limits_{i=1}^3 \Omega_i,\\
&& \label{U3x}\Omega_i=\frac{2c_i}{16\pi B_i}\left(\frac{s}{s_0}{\rm e}^{-\imath\frac{\pi}{2}} \right)^{\alpha_{IP_i}(0)-1}\exp\left[ -\frac{b^2}{4B_i}\right],\\
&& \label{U4}B_i=\alpha^{\prime}_{IP_i}\ln\left(\frac{s}{s_0}{\rm e}^{-\imath\frac{\pi}{2}} \right)+\frac{r_i^2}{4}.
\end{eqnarray}
The values of parameters $c_i$ and $r^2_i$ are derived in~(\ref{3IPtrajectories}) and listed 
in Table~\ref{tab:3IP}. 
Figs.~\ref{fig:2} demonstrate function $S(s/s_0,\xi,q_t)$ calculated for 
two values of energies a) $\sqrt{s}=62.7$~GeV and b) $\sqrt{s}=10$~TeV for different 
$\xi$ values: $\xi=0.3$ (dotted), $\xi=0.1$ (dashed) and $\xi=10^{-4}$ (solid).

\begin{center}
\begin{table}
\begin{center}
\begin{tabular}{|c|c|c|c|}
\hline
 $i$        &      1   & 2   & 3  \\
\hline
 $c_i$          &    $53.0\pm 0.8$   &   $9.68\pm 0.16$  &   $1.67\pm 0.07$ \\
\hline
  $r^2_i$ (GeV$^{-2}$)&     $6.3096\pm 0.2522$   &    $3.1097\pm 0.1817$ &   $2.4771\pm 0.0964$\\
\hline
\end{tabular}
\caption{\label{tab:3IP}Parameters of the model.}
\end{center}
\end{table}  
\end{center}

\subsubsection{Parametrization of $\pi^+ p$ cross section}
\label{sss:parametrization}

In the present version of generator we use 4 parametrizations for $\pi^+ p$ cross section.
\\

\noindent
{\bf The Donnachie-Landshoff (DL) parametrization~\cite{landshofftot}:}
\begin{equation}
\label{sigland}
\sigma^{tot}_{\pi^+p}(s)= 13.63\; s^{0.0808}+25.56\; s^{-0.4525},\; ({\rm mb}).
\end{equation} 

\noindent
{\bf The COMPETE parametrization~\cite{COMPETE}:} 
\begin{eqnarray}\label{sigcomplete}
&&\sigma^{tot}_{\pi^+p}(s)=Z_{\pi p}+B\ln^2\left(\frac{s}{s_0}\right)+
\left(Y_+s^{\alpha_+}-Y_-s^{\alpha_-}\right)/s ,\; ({\rm mb}).\\
&&Z_{\pi p}=21.23\pm 0.33\; {\rm mb},\;\nonumber \\ 
&&B=0.3152\pm 0.0095\; {\rm mb},\;\nonumber \\ 
&&s_0=34\pm 5.4\; {\rm GeV}^2,\;\nonumber \\
&&Y_+=17.8\pm 1.10,\; \alpha_+=0.533\pm 0.015,\;\nonumber \\
&&Y_-=5.72\pm 0.16,\; \alpha_-=0.4602\pm 0.0064.\nonumber
\end{eqnarray}

In the next two parametrizations total cross-section can be obtained thorough the optical theorem
\begin{equation}
\label{opticaltheorem}
\sigma^{tot}_{\pi^+p}=\frac{1}{s}\Im m \left. T(s,t)\right|_{t=0}.
\end{equation}

\noindent
{\bf The Bourrely-Soffer-Wu (BSW) parametrization~\cite{BSW}:} 
\begin{eqnarray}
\label{BSW0} 
&&{\rm T}(s,t_p)=\imath\int\limits_0^{\infty}b\; db\; J_0(b\sqrt{-t_p})(1-{\rm e}^{-\Omega_0(s,b)}),\\
&&\label{BSWomega} \Omega_0(s,b)=\Omega_{IP}+\sum\limits_{i}\Omega_i,\\
&&\label{BSWFFF}\Omega_{IP}\simeq \frac{s^c}{\ln^{c^{\prime}}s}\left[1+
\frac{{\rm e}^{\imath\pi c}}{\left( 1+\frac{\imath\pi}{\ln s}\right)^{c^{\prime}}} \right] F_{BSW}(b)\;
 {\rm for}\; s\gg m_p^2,|t|.
\end{eqnarray}

\noindent
For the $\pi^+ p$  we have $i=\rho$ in~(\ref{BSWomega}) and
\begin{eqnarray}
&& F^{\pi^+ p}_{BSW}(b)=\int\limits_0^{\infty}q\; dq\; J_0(q b) f_{\pi}\frac{a_{\pi}^2-q^2}{a_{\pi}^2+q^2}
\times\frac{1}{(1+\frac{q^2}{m_1^2})(1+\frac{q^2}{m_2^2})(1+\frac{q^2}{m_{3\pi}^2})},\\
&& \Omega_{\rho}\simeq C_{\rho}(1+\imath)\left( \frac{s}{s_0}\right)^{\alpha_{\rho}(0)-1}
\frac{{\rm e}^{-\frac{b^2}{4B_{\rho}}}}{2B_{\rho}},\\
&& B_{\rho}=b_{\rho}+\alpha_{\rho}^{\prime}(0)\ln\frac{s}{s_0},\; b_{\rho}=4.2704,\;\\ 
&& \alpha_{\rho}(t)=0.3202+t,\; C_{\rho}=4.1624,
\end{eqnarray}
where values of parameters are listed in Table~\ref{tab:BSW}.

\begin{center}
\begin{table}[t!]
\begin{center}
\begin{tabular}{|l|l|l|l|l|l|l|l|l|}
\hline
 $c$   &  $c^{\prime}$   &  $m_1$   & $m_2$ & $m_{3\pi}$ & $f_{\pi}$ & $a_{\pi}$ & $f$ & $a$   \\
\hline
 0.167 &  0.748  & 0.577225  &  1.719896 & 0.7665 & 4.2414 & 2.3272 & 6.970913 & 1.858442  \\
\hline
\end{tabular}
\end{center}
\caption{\label{tab:BSW}Parameters of the model~\cite{BSW}.}
\end{table}
\end{center}

\noindent
{\bf The Godizov-Petrov (GP) parametrization~\cite{godizov1},\cite{godizov2}.} \\

\noindent
In this parametrization  the scattering amplitude is represented in the usual
eikonal form
\begin{equation}
\label{eikna}
T(s,b) = \frac{e^{2i\delta(s,b)}-1}{2i}
\end{equation}
(here $T(s,b)$ is the amplitude in the impact parameter 
$b$ space, $s$ is the invariant mass squared of colliding particles and
$\delta(s,b)$ is the eikonal function). Amplitudes in
the impact parameter space and momentum one are related thorough
the Fourier-Bessel transforms 
\begin{eqnarray}
&&\label{fourbess} f(s,b) = \frac{1}{16\pi s}\int_0^{\infty}d(-t)J_0(b\sqrt{-t})f(s,t)\,,\\
&& f(s,t) = 4\pi s\int_0^{\infty}db^2J_0(b\sqrt{-t})f(s,b)\,.
\end{eqnarray}

\noindent
Eikonal function in the momentum space is
\begin{eqnarray}\label{eikmin}
&&\delta(s,t) = \delta_{\rm P}(s,t)+\delta_f(s,t)=\nonumber\\
&&
=\left(i+{\rm tg}\frac{\pi(\alpha_{\rm P}(t)-1)}{2}\right)
\beta_{\rm P}(t)\left(\frac{s}{s_0}\right)^{\alpha_{\rm P}(t)} + \nonumber \\
&&
+\left(i+{\rm tg}\frac{\pi(\alpha_f(t)-1)}{2}\right)
\beta_f(t)\left(\frac{s}{s_0}\right)^{\alpha_f(t)}.
\end{eqnarray}

\noindent
The parametrization for the pomeron residue is 
\begin{equation}
\label{respom}
\beta_{\rm P}(t) = B_{\rm P}e^{b_{\rm P}\,t}
(1+d_1\,t+d_2\,t^2+d_3\,t^3+d_4\,t^4)\,,
\end{equation}
which is approximately (at low values of 
$d_1$, $d_2$, $d_3$ è $d_4$) an exponential at low $t$ values. Residues
of secondary reggeons we set as exponentials:
\begin{equation}
\label{ressec}
\beta_f(t) = B_fe^{b_f\,t}.
\end{equation}

\begin{table}[ht!]
\begin{tabular}{|l|l|l|l|l|l|}
\hline
{\bf Pomeron} & & {\bf $f_2$-reggeon} & & {\bf $\omega$-reggeon} &  \\
\hline
$p_1$ & $0.123$ & $c_f$ & $0.1$ GeV$^2$ & $c_\omega$ & $0.9$ GeV$^2$ \\
$p_2$ & $1.58$ GeV$^{-2}$ & & & & \\
$p_3$ & $0.15$  & & & & \\
$B_{\rm P}$ & $43.5$ &  $B_f$  & $153$ & $B_\omega$ & $46$ \\
$b_{\rm P}$ & $2.4$ GeV$^{-2}$ & $b_f$  & $4.7$ GeV$^{-2}$ & $b_\omega$ & $5.6$ GeV$^{-2}$\\
$d_1$ & $0.43$ GeV$^{-2}$ & & & &   \\
$d_2$ & $0.39$ GeV$^{-4}$  & & & &   \\
$d_3$ & $0.051$ GeV$^{-6}$  & & & &   \\
$d_4$ & $0.035$ GeV$^{-8}$  & & & & \\
\hline
$\alpha_{\rm P}(0)$ & $1.123$ & $\alpha_f(0)$ & $0.78$ & $\alpha_\omega(0)$ & $0.64$ \\
$\alpha'_{\rm P}(0)$ & $0.28$ GeV$^{-2}$  & $\alpha'_f(0)$ & $0.63$ GeV$^{-2}$ 
& $\alpha'_\omega(0)$ & $0.07$ GeV$^{-2}$ \\
\hline
\end{tabular}
\caption{\label{gtab1} 
Values of parameters of the model~\cite{godizov1},\cite{godizov2} for $pp$ scattering.}
\end{table}

\noindent
Phenomenological parametrization for the "soft" po\-me\-ron trajectory is set to
\begin{equation}
\label{pomeron}
\alpha_{\rm P}(t) = 1+p_1\left[1-p_2\,t\left({\rm arctg}(p_3-p_2\,t)
                             -\frac{\pi}{2}\right)\right]\,.
\end{equation}
Trajectories of secondary reggeons $f_2$ and $\omega$ are parametrized by
functions
\begin{equation}
\label{second}
\alpha_{\rm R}(t) = \left(\frac{8}{3\pi}
\alpha_s(\sqrt{-t+c_{\rm R}})\right)^{1/2},\; {\rm R}=f,\omega,
\end{equation}
where
\begin{equation}
\label{analytic}
\alpha_s(\mu) = \frac{4\pi}{11-\frac{2}{3}n_f}
\left(\frac{1}{\ln\frac{\mu^2}{\Lambda^2}}
+\frac{1}{1-\frac{\mu^2}{\Lambda^2}}\right)
\end{equation}
is the one-loop analytic QCD running coupling~\cite{solovtsov}, $n_f = 3$ is the number of flavours, $\Lambda\equiv\Lambda^{(3)} = 0.346$~GeV~\cite{bethke}. Parameters 
$c_f\,,\;c_\omega>0$ are rather small to spoil the asymptotic behaviour
of secondary trajectories in the perturbative domain.
Residues for $\pi\pi$, $\pi p$ and $pp$  are assumed to be
\begin{equation}
\label{relproc}
\beta^{\pi\pi}_P(t)=\frac{\beta^{\pi p}_P(t)\beta^{\pi p}_P(t)}{\beta^{pp}_P(t)},
\end{equation}
\begin{equation}
\label{relprocxx}
\beta^{\pi\pi}_f(t)=\frac{\beta^{\pi p}_f(t)\beta^{\pi p}_f(t)}{\beta^{pp}_f(t)}.
\end{equation}
Parameters of the model are listed in Tables~\ref{gtab1}.

\subsection{Double Pion Exchange}
\label{ss:DPE}
The diagram of the Double Pion Exchange (D$\pi$E) process $p+p\to n+X+n$ is presented in 
Fig.~\ref{fig:1}b.  The momenta are $p_1$, $p_2$, $p_{n_1}$, $p_X$, $p_{n_2}$ respectively. 
In the center-of-mass frame these can be represented as follows
\begin{eqnarray}
&& \label{kindpie1}p_{\pi_i}\simeq\left( \xi_i\frac{\sqrt{s}}{2},(-1)^{i-1}\xi_i\frac{\sqrt{s}}{2},\mbox{\bf q}_i\right),\\
&& \label{kin1dpie2}p_{n\; i}=p_i-p_{\pi_i},\\
&& p_X^2=M^2=\xi_1\xi_2 s \beta^2\frac{1+\beta^2}{2}-\left(\mbox{\bf q}_1+\mbox{\bf q}_2\right)^2-
m_p^2\beta^2(\xi_1^2+\xi_2^2)+\nonumber\\
&&+(t_1+t_2+2(m_p^2-m_n^2))
\cdot\left(\beta^2 (\xi_1+\xi_2)+\frac{t_1+t_2+2(m_p^2-m_n^2)}{s} \right)
\label{kindpie5}\simeq\xi_1\xi_2 s,\\
&&-t_i\simeq\frac{\mbox{\bf q}^2_i+\xi_i^2m_p^2}{1-\xi_i}\label{kindpie6}.
\end{eqnarray}
 
The cross-section can be evaluated as follows:
\begin{eqnarray}
&& \label{Udsigmad}d\sigma=S_2(s/s_0,\xi_{1,2},\mbox{\bf q}^2_{1,2}) d\sigma_0,\\
&&\label{dsigma0qtd} \frac{d\sigma_0(\xi_1,\xi_2,\mbox{\bf q}^2_1,\mbox{\bf q}^2_2)}
{d\xi_1 d\xi_2 d\mbox{\bf q}^2_1 d\mbox{\bf q}^2_2}=
\prod\limits_{i=1}^2 \left[ (m_p^2\xi_i^2+\mbox{\bf q}^2_i)|\Phi_B(\xi_i,\mbox{\bf q}^2_i)|^2 
\frac{\xi_i}{(1-\xi_i)^2}\right]\cdot\;\sigma_{\pi^+\pi^+}(\xi_1\xi_2 s),\\
&& \label{survivald}S_2=\frac{\sum\limits_{i,j=0,s} \rho_{ij}^2|\bar{\Phi}_{ij}(s/s_0,
\xi_{1,2},\mbox{\bf q}^2_{1,2})|^2}{\prod\limits_{i=1}^2\left[(m_p^2\xi_i^2+\mbox{\bf q}^2_i)|
\Phi_B(\xi_i,\mbox{\bf q}^2_i)|^2\right]},\\
&& \label{fiij} \bar{\Phi}_{ij}=\frac{N(\xi_1)N(\xi_2)}{(2\pi)^2}\cdot
\int\limits_0^{\infty} db_1 db_2 \Theta_i(b_1,\xi_1,|\mbox{\bf q}_1|)
\Theta_j(b_2,\xi_2,|\mbox{\bf q}_2|) I_{\phi}(b_1,b_2),\\
&& I_{\phi}(b_1,b_2)=\int\limits_0^{\pi}\frac{d\phi}{\pi} V \left(\sqrt{b_1^2+b_2^2-2b_1b_2\cos\phi}\right),\\
&& \rho_{00}=m_p^2\xi_1\xi_2,\quad \rho_{0s}=m_p\xi_1,\quad \rho_{s0}=m_p\xi_2,\quad \rho_{ss}=1.
\end{eqnarray}
For low $t_i$ the function $S_2$ is approximately equal to
\begin{eqnarray}
&&\label{survDpiEapprox} F(\xi_1,\xi_2)\equiv S_2(s/s_0,\xi_1,\xi_2,0,0)\simeq\nonumber\\
&& \simeq \left( \sqrt{S(s/s_0,\xi_1,0)}+\sqrt{S(s/s_0,\xi_2,0)}\right.-
\left.\sqrt{S(s/s_0,\xi_1,0)S(s/s_0,\xi_1,0)} \right)^2.
\end{eqnarray}

\begin{figure}[h!]
\begin{center} 
\includegraphics[width=\textwidth]{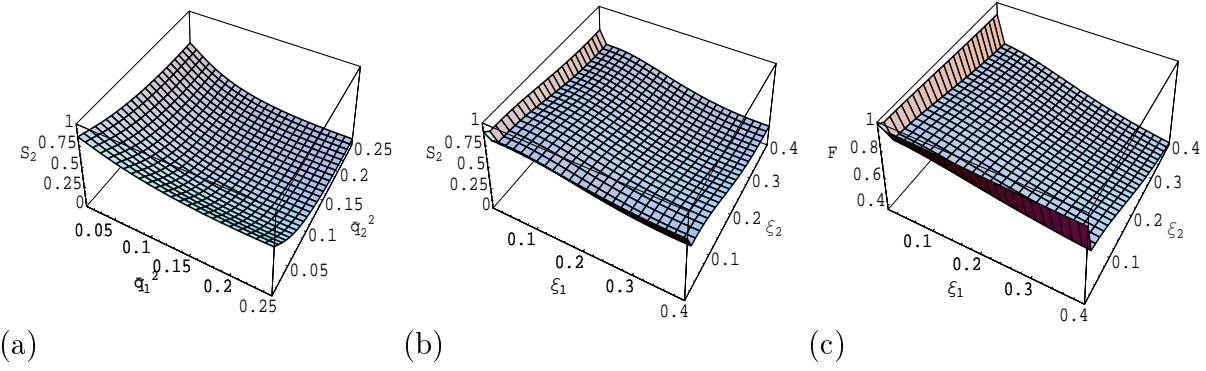}
\caption{\label{fig:3}The function $S_2(s/s_0,\xi_{1,2},|\mbox{\bf q}_{1,2}|)$ at  $\sqrt{s}=10$~TeV: 
a) for fixed $\xi_{1,2}=0.01$ b) for fixed $|\mbox{\bf q}_{1,2}|\sim 0$. 
c) The function $F(\xi_1,\xi_2)$ at $\sqrt{s}=10$~TeV.}
\end{center} 
\end{figure} 
\noindent
Figs.~\ref{fig:3} demonstrates 2D projections of function $S_2(s/s_0,\xi_{1,2},|\mbox{\bf q}_{1,2}|)$ 
and function $F(\xi_1,\xi_2)$ at $\sqrt{s}=10$~TeV. 

To obtain $\pi^+\pi^+$ cross-sections we use parametrizations described in the
subsection~\ref{sss:parametrization} with the following approximations:
\begin{equation}
\sigma^{tot}_{\pi^+\pi^+}\simeq \frac{\left(\sigma^{tot}_{\pi^+p}\right)^2}{\sigma^{tot}_{pp}}
\end{equation}
for the {\bf DL} and {\bf COMPETE} ones, 
quantities
\begin{equation}
F^{\pi^+\pi^+}_{BSW}(b)\simeq\int\limits_0^{\infty}q\; dq\; J_0(q b)\; f_{\pi\pi}\frac{a_{\pi\pi}^2-q^2}{a_{\pi\pi}^2+q^2}\frac{1}{(1+\frac{q^2}{m_{3\pi}^2})^2},
\end{equation}
\begin{equation}
\Omega_0\simeq\Omega_{IP},\; f_{\pi\pi}=\frac{f_{\pi}^2}{f},\; \frac{1}{a_{\pi\pi}^2}=\frac{2}{a_{\pi}^2}-\frac{1}{a^2}.
\end{equation}
for the {\bf BSW} one, which should be substituted to~(\ref{BSWFFF}), and
equations~(\ref{relproc}),(\ref{relprocxx}) for the {\bf GP} one.

\subsection{Relative contributions of $\pi$, $\rho$ and $a_2$ reggeons.}
\label{ss:relative}
  
For $\rho$ and $a_2$ contributions formulae are similar to ones
described in the chapters (\ref{ss:SPE}) and (\ref{ss:DPE}).
\begin{eqnarray}
&&\label{csSRE}\frac{d\sigma_{\rm S\reg E}}{d\xi dt}=F_{\reg}(\xi, t)S_{\reg}(s/s_0, 
\xi, t)\;\sigma_{\reg^+ p}(\xi s) ,\\
&&\label{csDRpiE} \frac{d\sigma_{{\rm D\reg}\pi {\rm E}}}{d\xi_1d\xi_2dt_1dt_2}
=F_{\reg\pi}(\xi_1, \xi_2,t_1, t_2)S_{\reg,2}(s/s_0, \{\xi_i\},\{t_i\})
\times\sigma_{\reg^+\pi^+}(\xi_1\xi_2 s),\\
&&\label{FRform} F_{\reg}(\xi,t)=\frac{|\eta_{\reg}|^2\tilde{G}_{\reg^+pn}^2}
{16\pi^2}{\rm e}^{2b_Rt}\xi^{1-2\alpha_{\reg}(t)}\left(1+\kappa_{\reg}^2\frac{\mbox{\bf q}^{\;2}}{4m_p^2}\right),\\
&&F_{\reg\pi}(\{\xi_i\},\{t_i\})=F_0(1)F_{\reg}(2)+F_0(2)F_{\reg}(1)+\nonumber\\
&&+2\sqrt{\frac{F_0(1)F_0(2)F_{\reg}(1)F_{\reg}(2)}{t_1t_2(1-\xi_1)(1-\xi_2)}}
\times \frac{\left(m_p\xi_1+\mbox{\bf q}_{1}^{\;2}\frac{\kappa_{\reg}}{2m_p}\right)\left(m_p\xi_2+\mbox{\bf q}_{2}^{\;2}
\frac{\kappa_{\reg}}{2m_p}\right)}{\left(1+\mbox{\bf q}_{1}^{\;2}
\frac{\kappa_{\reg}^2}{4m_p^2}\right)\left(1+\mbox{\bf q}_{2}^{\;2}\frac{\kappa_{\reg}^2}{4m_p^2}\right)},\\
&&F_{0,\reg}(i)= F_{0,\reg}(\xi_i,t_i),\; \mbox{\bf q}_i^{\;2}\simeq-t_i(1-\xi_i)-m_p^2\xi_i^2.
\end{eqnarray}
Here $\kappa_{\reg}=8$ is the ratio of spin-flip to nonflip amplitude, $\alpha_{\reg}(t)\simeq 0.5+0.9t$ 
and parameters for $\rho$, $a_2$ mesons are~\cite{rhoa2pars}
\begin{eqnarray}
&& \eta_{\rho}=-\imath+1,\; \eta_{a_2}=\imath+1,\\
&& b_{\rho}=2\;{\rm GeV}^{-2},\; b_{a_2}=1\;{\rm GeV}^{-2},\\
&& \frac{\tilde{G}_{\rho^+pn}^2}{8\pi}=0.18\;{\rm GeV}^{-2},\; 
\frac{\tilde{G}_{{a_2}^+pn}^2}{8\pi}=0.405\;{\rm GeV}^{-2}.
\end{eqnarray}
Rescattering corrections $S_{\reg}$ and $S_{\reg,2}$ are calculated by the method used 
in~\cite{ourneutrontot},\cite{ourneutronel}.
Basic assumptions in our calculations are: 
\begin{itemize}
\item $\rho\;\rho$, $\rho\; a_2$ and $a_2\; a_2$ contributions are small; 
\item interference terms of the type $T^*_{{\rm S}\pi{\rm E}}T_{{\rm S\reg E}}$, 
$T^*_{{\rm D\reg}\pi{\rm E}}T_{{\rm D\reg}^{\prime}\pi{\rm E}}$ are small~\cite{KMRn1c16}, 
${\rm \reg, \reg}^{\prime}=\pi,\;\rho,\;a_2$, $\reg\neq \reg^{\prime}$, 
where $T$ are amplitudes of the corresponding processes;
\item approximate relations $\sigma_{\reg^+ p}\simeq\sigma_{\pi^+ p}$, 
$\sigma_{\reg^+\pi^+}\simeq\sigma_{\pi^+\pi^+}$~\cite{KMRn1c16}.
\end{itemize}

\begin{figure}[h!]
\begin{center} 
\includegraphics[width=\textwidth]{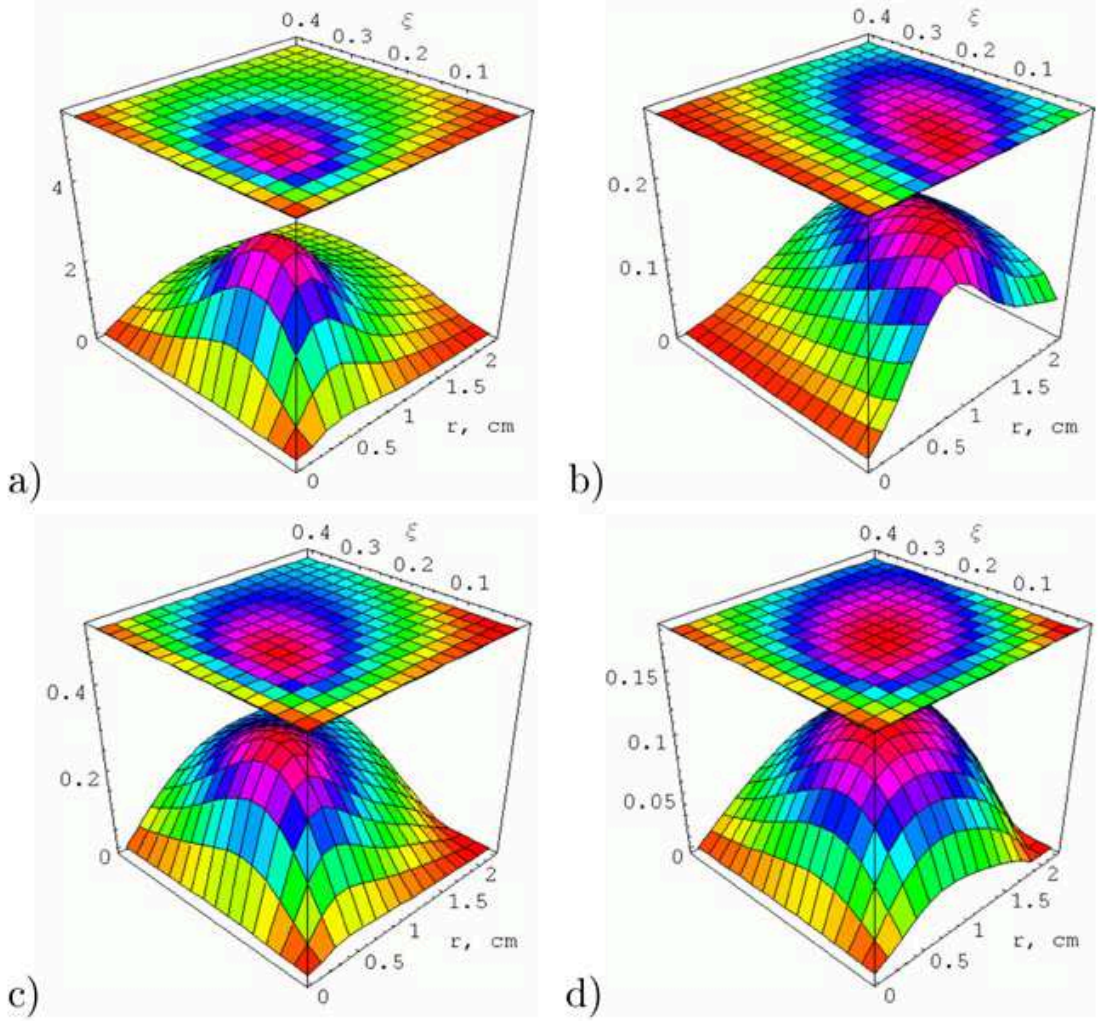}
\caption{\label{fig:4} Cross-sections $\frac{d\sigma}{d\xi dr}$ in $mb\cdot cm^{-1}$ at 
$\sqrt{s}=7$~TeV for: a) S$\pi$E; b) S$\rho$E+S$a_2$E; c) D$\pi$E; d) D$\rho\pi$E+D$a_2\pi$E. $r$ is the transverse distance from
the beam.}
\end{center} 
\end{figure}

\noindent
Figs.~\ref{fig:4} demonstrates 3D plots for cross sections of the Single and Double Reggeon
Exchange reactions for the different reggeons
(S$\pi$E a), S$\rho$E+S$a_2$E b), D$\pi$E c) and D$\rho\pi$E+D$a_2\pi$E d))   
at $\sqrt{s}=7$~TeV. 

\newpage

\section{Program Overview}
\label{s:programoverview}

The kinematics of S$\reg$E and D$\reg$E reactions, 
\begin{equation}\label{sre}
pp\to n+(\reg p)\to n+X
\end{equation}
and
\begin{equation}\label{dre}
pp\to n+(\reg\reg)+n\to n+X+n,
\end{equation}
are defined by 
the relative energy loss  of  $\xi_n$ and the square of the transverse momentum $t_n$ of the 
leading neutron. The vertex $p\reg n$ is generated on the basis of the 
models described above. The differential cross sections for the generated neutron and reggeon
are calculated according the selected models for absorptive corrections and for $\reg p$ ($\reg\reg$)
interactions. Then,  PYTHIA 6.420 \cite{pythia} is called for the 
$\reg p\to X$ generation in the case of S$\reg$E and $\reg\reg \to X$ 
generation in the case of D$\reg$E. Parameters of the all generated particles, including 
beam protons,  leading neutrons, reggeons and X, products of $\reg p$ ($\reg\reg$) interaction, 
are stored in \Py common blocks.

\subsection{Main Subroutines}
\label{ss:mainsubroutines}

\drawbox{SUBROUTINE MONINIT}\label{p:MONINIT}
\begin{entry}
\itemc{Purpose:} to initialize the generation procedure. In particulary, 
\begin{subentry}
\item[-] to show program title;
\item[-] to read control parameters from the file \ttt{moncher.par};
\item[-] to set default \Mo\ parameters;
\item[-] to initilize \Py;
\item[-] to initilize LHE format output.
\end{subentry}
\iteme{Status of call:} should be called obligatory, one time, in the begining of the
main program before calling of \ttt{MONEVEN}.
\iteme{Calling by:} main program
\iteme{Calling of:} \ttt{MONTITL}, \ttt{MONPARA}, \ttt{MONMBDF}, 
                    \ttt{MONUPIN}, \ttt{PYINIT}
\end{entry}

\drawbox{SUBROUTINE MONTITL}\label{p:MONTITL}
\begin{entry}
\itemc{Purpose:} to print title of \Mo\ on the screen. Namely,  
\begin{tabbing}      
{\tt *********************************}\\[-1.1mm]
{\tt *\ \ \ \ \ \ \ \ \ \ \ \ \ \ \ \ \ \ \ \ \ \ \ \ \ \ \ \ \ \ \ *}\\[-1.1mm]
{\tt *\ \ MON-te-carlo generator for\ \ \                           *}\\[-1.1mm]
{\tt *\ \ \ \ \ CH-arge\ \ \ \ \ \ \ \ \ \ \ \ \ \ \ \ \ \ \        *}\\[-1.1mm]
{\tt *\ \ \ \ \ \ \ E-xchange\ \ \ \ \ \ \ \ \ \ \ \ \ \ \          *}\\[-1.1mm]
{\tt *\ \ \ \ \ \ \ \ R-eactions\ \ \ \ \ \ \ \ \ \ \ \ \           *}\\[-1.1mm]
{\tt *\ \ \ \ \ \ \ \ \ \ \ \ \ \ \ \ \ \ \ \ \ \ \ \ \ \ \ \ \ \ \ *}\\[-1.1mm]
{\tt *\ \ Version 1.1.0.(12/03/2011)\ \ \                           *}\\[-1.1mm]
{\tt *\ \ \ \ \ \ \ \ \ \ \ \ \ \ \ \ \ \ \ \ \ \ \ \ \ \ \ \ \ \ \ *}\\[-1.1mm]
{\tt *\ \ \ \ \ \ \ \ \ \ \ \ \ \ \ \ \ \ \ \ \ \ \ \ \ \ \ \ \ \ \ *}\\[-1.1mm]
{\tt *\ \ \ \ \ \ \ \ \ \ \ \ \ \ \ \ \ \ \ \ \ \ \ \ \ \ \ \ \ \ \ *}\\[-1.1mm]      
{\tt *\ \ R.Ryutin,A.Sobol,V.Petrov\ \ \ \                          *}\\[-1.1mm]
{\tt *\ \ \ \ \ \ \ \ \ \ \ \ (IHEP,Protvino)\ \ \ \                *}\\[-1.1mm]
{\tt *\ \ \ \ \ \ \ \ \ \ \ \ \ \ \ \ \ \ \ \ \ \ \ \ \ \ \ \ \ \ \ *}\\[-1.1mm]
{\tt *********************************}\\[-1.1mm]
\end{tabbing}
\iteme{Status of call:} should be called OBLIGATORY.
\iteme{Calling by:} \ttt{MONINIT}
\end{entry}

\newpage
\drawbox{SUBROUTINE MONPARA}\label{p:MONPARA}
\begin{entry}
\itemc{Purpose:} to read control parameters for the \Mo\ 
and \Py\ generation from the file \ttt{moncher.par} 
Control parameters for \Mo\ are called \ttt{MONPAR}, 
they are stored to the common block \ttt{/MONGLPA/}. 
Any \Py parameters can be defined  for \Py common blocks\\
\ttt{/PYJETS/,/PYDAT1/,/PYDAT2/,/PYDAT3/,/PYDAT4/,/PYDATR/,/PYSUBS/,
     /PYPARS/,/PYINT1/,/PYINT2/,/PYINT3/,/PYINT4/,/PYINT5/,/PYINT6/,
     /PYINT7/,/PYINT8/,/PYMSSM/,/PYMSRV/,/PYTCSM/,/PYPUED/} 
(see \cite{pythia}).
\iteme{Status of call:} can be called if you like to define some control 
parameters from the \ttt{moncher.par}. By default, \Mo\ initilizes a generation
of minimum bias events by \Py\ with some default parameters. 
\iteme{Calling by:} \ttt{MONINIT}
\iteme{Calling of:} \ttt{MONGIVE}
\end{entry}

\drawbox{SUBROUTINE MONMBDF}\label{p:MONMBDF}
\begin{entry}
\itemc{Purpose:} to define default parameters for the generation.
By default, \Mo\ and \Py\ parameters are defined to
generate 10 minimum bias events at c.m.s. energy 7 TeV. 
\iteme{Status of call:} to be called at initialization. Default parameters are 
redefined by the call of the \ttt{MONPARA} reading parameters from  the file 
\ttt{moncher.par}.   
\iteme{Calling by:} \ttt{MONINIT}
\iteme{Calling of:} \ttt{MONGIVE}
\end{entry}

\drawbox{SUBROUTINE MONEVEN}\label{p:MONEVEN}
\begin{entry}
\itemc{Purpose:} call subroutines for the single event generation 
\begin{subentry}
\iteme{MONPAR(7)= 1 :} call \ttt{MONSPEG} for Single Charge Exchange (SCE) generation.
\iteme{MONPAR(8)= 1 :} call \ttt{MONDPEG} for Double Charge Exchange (DCE) generation.
\iteme{MONPAR(7)= 0 and MONPAR(8)= 0 :} call \ttt{PYEVNT} for the \Py event 
generation.
\end{subentry}
\iteme{Status of call:} should be be called in the user main program, 
in the cycle of events.
\iteme{Calling by:}  main program
\iteme{Calling of:} \ttt{MONSPEG}, \ttt{MONDPEG}, \ttt{PYEVNT}
\end{entry}

\drawbox{SUBROUTINE MONSPEG}\label{p:MONSPEG}
\begin{entry}
\itemc{Purpose:} to generate single SCE event, 
$p_{1}^{beam}p_{2}^{beam}\to n (\pi_{virt}^+ p_{2}^{beam}) \to n X$, 
in the following sequence:
\begin{subentry}
\item{-} the vertex $p_{1}^{beam}n\pi_{virt}^+$ is generated by \ttt{MONSPEM};
\item{-} \Py\ is initialized for the generation of $\pi_{virt}^+ p_{2}^{beam}$ interaction;
\item{-} \Py\ is called for the generation, hadronization and decays;
\item{-} the \Py output is rewriting to include beam protons and neutron to the
final state of the reaction with the particles from $X$.
\end{subentry}
Simulation of $\pi_{virt}^+ p_{2}^{beam}$ interaction is 
controled  by \Py parameters. It can be elastic, minimum bias
or diffractive interaction. Number of the corresponding SCE process is equal
to the number of the \Py process + 500.
\iteme{Status of call:} called if \ttt{MONPAR(7)=1}. 
\iteme{Calling by:} \ttt{MONEVEN}
\iteme{Calling of:} \ttt{MONSPEM}, \ttt{MONSHPY}, \ttt{PYINIT}, \ttt{PY1ENT}, 
                    \ttt{PYANGL}
\end{entry}

\drawbox{SUBROUTINE MONDPEG}\label{p:MONDPEG}
\begin{entry}
\itemc{Purpose:} to generate single DCE event, 
$p_{1}^{beam}p_{2}^{beam}\to n (\pi_{virt}^+\pi_{virt}^+) n \to n X n$, 
in the following sequence:
\begin{subentry}
\item{-} the vertexes $p_{1}^{beam}n\pi_{virt}^+$ and $p_{2}^{beam}n\pi_{virt}^+$ 
are  generated by \ttt{MONDPEM};
\item{-} \Py\ is initialized for the generation of $\pi_{virt}^+ \pi_{virt}^+$ 
interaction;
\item{-} \Py\ is called for the generation, hadronization and decays;
\item{-} the \Py output is rewriting to include beam protons and neutrons to the
final state of the reaction with the particles from  $X$.
\end{subentry}
Simulation of $\pi_{virt}^+ \pi_{virt}^+$ interaction is 
controled  by \Py parameters. It can be elastic, minimum bias
or diffractive interaction. Number of the corresponding DCE process is equal
to the number of the \Py process + 600.
\iteme{Status of call:} called if \ttt{MONPAR(8)=1}. 
\iteme{Calling by:} \ttt{MONEVEN}
\iteme{Calling of:} \ttt{MONDPEM}, \ttt{MONSHPY}, \ttt{PYINIT}, \ttt{PY1ENT}, 
                    \ttt{PYANGL}
\end{entry}

\drawbox{SUBROUTINE MONSPEM(NO,PN,PR,M2)}\label{p:MONSPEM}
\begin{entry}
\itemc{Purpose:}  to generate momentums and energies of neutron $n$ and
virtual exchange reggeon $\reg^+$ in the reaction of Single Charge Exchange:\\
$p_{1}^{beam}p_{2}^{beam}\to n (\reg^+ p_{2}^{beam})  \to n X $ 

\iteme{INTEGER NO ({\it input}) :}  type of exchange reggeon $\reg$;
\begin{subentry}
\iteme{= 1 :} $\pi^+$ 
\iteme{= 2 :} $\rho^+$ 
\iteme{= 3 :} $a_2^+$ 
\end{subentry}

\iteme{DOUBLE PRECISION PN(5) ({\it output})  :}\label{p:PN}  
kinematical parameters of  the neutron $n$.
\begin{subentry}
\iteme{PN(1) :} $p_x$, momentum of neutron in the $x$ direction, in GeV/$c$.
\iteme{PN(2) :} $p_y$, momentum of neutron in the $y$ direction, in GeV/$c$.
\iteme{PN(3) :} $p_z$, momentum of neutron in the $z$ direction, in GeV/$c$.
\iteme{PN(4) :} $E$, energy of neutron, in GeV.
\iteme{PN(5) :} $m$, mass of neutron, in GeV/$c^2$. 
\end{subentry}

\iteme{DOUBLE PRECISION PR(5) ({\it output})  :}\label{p:PR} 
kinematical parameters of the reggeon $\reg^+$. 
\begin{subentry}
\iteme{PR(1) :} $p_x$, momentum of reggeon in the $x$ direction, in GeV/$c$.
\iteme{PR(2) :} $p_y$, momentum of reggeon in the $y$ direction, in GeV/$c$.
\iteme{PR(3) :} $p_z$, momentum of reggeon in the $z$ direction, in GeV/$c$.
\iteme{PR(4) :} $E$, energy of reggeon, in GeV.
\iteme{PR(5) :} $m$, mass of reggeon, in GeV/$c^2$. 
\end{subentry}

\iteme{DOUBLE PRECISION M2 ({\it output})     :}\label{p:M2CE} 
invariant mass of the system $(\reg^+p^{beam2})$, in GeV/$c^2$.

\iteme{Calling by:} \ttt{MONSPEG}
\iteme{Calling of:} \ttt{MONGE2D}
\end{entry}

\drawbox{SUBROUTINE MONDPEM(NO,PN1,PN2,PR1,PR2,M2)}\label{p:MONDPEM}
\begin{entry}
\itemc{Purpose:} to generate momentums and energies of neutrons $n$ and
virtual exchange reggeons $\reg^+$ in the reaction of Double Charge Exchange:\\
$p_{1}^{beam}p_{2}^{beam}\to n (\reg_{1}^+\reg_{2}^+) n \to n X n$

\iteme{INTEGER NO ({\it input}) :}  type of exchange reggeons $\reg_{1}^+\reg_{2}^+$;
\begin{subentry}
\iteme{= 1 :} $\pi^+\pi^+$ 
\iteme{= 2 :} $\pi^+\rho^+$ 
\iteme{= 3 :} $\pi^+a_2^+$ 
\end{subentry}

\iteme{DOUBLE PRECISION PN1(5),PN2(5) ({\it output})  :}\label{p:PN12}  
kinematical parameters of  the neutron~$n$.
\begin{subentry}
\iteme{PN1(1),PN2(1) :} $p_x$, momentum of neutrons in the $x$ direction, in GeV/$c$.
\iteme{PN1(2),PN2(2) :} $p_y$, momentum of neutrons in the $y$ direction, in GeV/$c$.
\iteme{PN1(3),PN2(3) :} $p_z$, momentum of neutrons in the $z$ direction, in GeV/$c$.
\iteme{PN1(4),PN2(4) :} $E$, energy of neutrons, in GeV.
\iteme{PN1(5),PN2(5) :} $m$, mass of neutrons, in GeV/$c^2$. 
\end{subentry}

\iteme{DOUBLE PRECISION PR1(5),PR2(5)({\it output})  :}\label{p:PR12} 
kinematical parameters of the reggeons~$\reg_{1,2}^+$. 
\begin{subentry}
\iteme{PR(1),PR2(1) :} $p_x$, momentum of reggeons in the $x$ direction, in GeV/$c$.
\iteme{PR(2),PR2(2) :} $p_y$, momentum of reggeons in the $y$ direction, in GeV/$c$.
\iteme{PR(3),PR2(3) :} $p_z$, momentum of reggeons in the $z$ direction, in GeV/$c$.
\iteme{PR(4),PR2(4) :} $E$, energy of reggeons, in GeV.
\iteme{PR(5),PR2(5) :} $m$, mass of reggeons, in GeV/$c^2$. 
\end{subentry}

\iteme{DOUBLE PRECISION M2 ({\it output})     :}\label{p:M2DCE} 
invariant mass of the system $(\reg_{1}^+\reg_{2}^+)$, in GeV/$c^2$.

\iteme{Calling by:} \ttt{MONDPEG}
\iteme{Calling of:} \ttt{MONGE2D}, \ttt{MONGE2D4}
\end{entry}

\drawbox{SUBROUTINE MONSHPY(NSHIFT)}\label{p:MONSHPY}
\begin{entry}
\itemc{Purpose:} to shift data of arrays of the \Py common block \ttt{/PYJETS/}
for \ttt{NSHIFT} positions. It should be done to fill first  \ttt{NSHIFT} positions
of \ttt{/PYJETS/} arrays by the parameters of the beam protons and neutrons in the
 final state of reaction.

\iteme{Calling by:} \ttt{MONSPEG}, \ttt{MONDPEG}
\end{entry}

\drawbox{SUBROUTINE MONGIVE(CHIN)}\label{p:MONGIVE}
\begin{entry}
\itemc{Purpose:} modification of the \Py subroutine  \ttt{PYGIVE} 
to set the value of any variable residing in the
commmonblocks \ttt{PYJETS}, \ttt{PYDAT1}, \ttt{PYDAT2}, \ttt{PYDAT3},
\ttt{PYDAT4}, \ttt{PYDATR}, \ttt{PYSUBS}, \ttt{PYPARS}, \ttt{PYINT1},
\ttt{PYINT2}, \ttt{PYINT3}, \ttt{PYINT4}, \ttt{PYINT5}, \ttt{PYINT6},
\ttt{PYINT7}, \ttt{PYINT8}, \ttt{PYMSSM}, \ttt{PYMSRV}, \ttt{PYTCSM}
or \ttt{MONGLPA}. 
This is done in a more controlled fashion than by directly including 
the common blocks in your program, in that array bounds are checked 
and the old and new values for the variable changed are written to 
the output for reference. In the following example,  
"\ttt{CALL MONGIVE('MONPAR(3)=14000')}", we have changed pp c.m.s. 
energy to 14 TeV. More detail explanation see in Ref.~\cite{pythia}
for subroutine \ttt{PYGIVE}.

\iteme{CHARACTER CHIN*(*) ({\it input}) :}\label{p:CHIN} 
character expression of length at most 100 characters, 
with requests for variables to be changed.  

\iteme{Calling by:} \ttt{MONPARA}, \ttt{MONMBDF}
\end{entry}

\drawbox{SUBROUTINE MONUPEV}\label{p:MONUPEV}
\begin{entry}
\itemc{Purpose:}  to write information about generated processes to the 
file \ttt{moncher.lhe} using special LHE record format.
For more detail information about LHE format see Ref.~\cite{LHE}.

\iteme{Status of call:} called if \ttt{MONPAR(2)=1}. 
\iteme{Calling by:} \ttt{MONINIT}
\end{entry}

\drawbox{SUBROUTINE MONUPIN}\label{p:MONUPIN}
\begin{entry}
\itemc{Purpose:}  to save information about all stable particles generated
in the event to the file \ttt{moncher.lhe} using special LHE record format.
For more detail information about LHE format see Ref.~\cite{LHE}.

\iteme{Status of call:} should be called for each generated event 
if \ttt{MONPAR(2)=1}. 
\iteme{Calling by:} user main program
\end{entry}

\subsection{Auxiliary Subroutines}
\label{ss:auxsubroutines}

These subroutines are used for internal calculations and should not be changed.

\drawbox{SUBROUTINE MONGE2D(FF,X1,X2,N1,N2,FF1,FF2,FF3,RG,XG,IG)}\label{p:MONGE2D}
\begin{entry}
\itemc{Purpose:} to generate two variables according to the 2D distribution from the table.

\iteme{DOUBLE PRECISION FF(N1,N2)({\it input})  :}\label{p:FFGE2D} \ttt{N1}$\times$\ttt{N2} dimensional 
 interpolation table of 2D distribution.
\iteme{DOUBLE PRECISION X1(N1),X2(N2))({\it input})  :}\label{p:X1X2GE2D} arrays of variables corresponding
to the table \ttt{FF}.
\iteme{INTEGER N1,N2({\it input})  :}\label{p:N1N2GE2D} dimensions of the 2D table.
\iteme{DOUBLE PRECISION FF1(N1),FF2(N1)({\it input})  :}\label{p:FF12GE2D} auxiliary integrated 
 tables for 2D distribution.
\iteme{DOUBLE PRECISION FF3(2,N1,N2)({\it input})  :}\label{p:FF3GE2D} auxiliary sums
from the table for 2D distribution.
\iteme{DOUBLE PRECISION RG(2)({\it input})  :}\label{p:RGGE2D} array for generated random numbers from 0 to 1.
\iteme{DOUBLE PRECISION XG(2)({\it output})  :}\label{p:XGGE2D} array for generated variables according to the 2D distribution.
\iteme{INTEGER IG(2)({\it output})  :}\label{p:IGGE2D} auxiliary numbers of the nearest to the \ttt{XG(2)} discrete point. 

\iteme{Calling by:} \ttt{MONSPEM}, \ttt{MONDPEM}
\end{entry}

\drawbox{SUBROUTINE MONG2D4(FF,X1,X2,N1,N2,FF1,FF2,FF3,II,XX,RG,XG)}\label{p:MONG2D4}
\begin{entry}
\itemc{Purpose:} to generate four variables according to the 4D distribution from the table.

\iteme{DOUBLE PRECISION FF(N1,N1,N2,N2)({\it input})  :}\label{p:FFG2D4}  \ttt{N1}$\times$\ttt{N1}$\times$\ttt{N2}$\times$\ttt{N2}   
dimensional interpolation table of 4D distribution.
\iteme{DOUBLE PRECISION X1(N1),X2(N2))({\it input})  :}\label{p:X1X2G2D4} arrays of variables corresponding
to the table \ttt{FF}.
\iteme{INTEGER N1,N2({\it input})  :}\label{p:N1N2G2D4} dimensions of the 4D table.
\iteme{DOUBLE PRECISION FF1(4,N1,N1,N2),FF2(4,N1,N1,N2)({\it input})  :}\label{p:FF12G2D4} auxiliary integrated 
 tables for the 4D distribution.
\iteme{DOUBLE PRECISION FF3(8,N1,N1,N2,N2)({\it input})  :}\label{p:FF3G2D4} auxiliary sums
from the table for the 4D distribution.
\iteme{INTEGER II(2)({\it input})  :}\label{p:IIG2D4} auxiliary numbers for multidimensional calculations.
\iteme{DOUBLE PRECISION XX(2)({\it input})  :}\label{p:XXG2D4} auxiliary points for multidimensional calculations.
\iteme{DOUBLE PRECISION RG(4)({\it input})  :}\label{p:RGG2D4} array for generated random numbers from 0 to 1.
\iteme{DOUBLE PRECISION XG(4)({\it output})  :}\label{p:XGG2D4} array for generated variables according to the 4D distribution.

\iteme{Calling by:} \ttt{MONDPEM}
\end{entry}

\drawbox{SUBROUTINE MONCUBI(FF,VS,FUN)}\label{p:MONCUBI}
\begin{entry}
\itemc{Purpose:} cubic spline interpolation for a function in the variable $\ln s$.

\iteme{DOUBLE PRECISION FF(6)({\it input})  :}\label{p:FFMONCUBI}  table of the function at six values of variable $s$ stored 
in the array XSQ(6) (see below the commonblock \ttt{MONTAB1}).
\iteme{DOUBLE PRECISION VS({\it input})  :}\label{p:VSMONCUBI} input value of $s$.
\iteme{DOUBLE PRECISION FUN({\it output})  :}\label{p:FUNMONCUBI} output value of the function.

\iteme{Calling by:} \ttt{MONDATA}
\end{entry}





\drawbox{SUBROUTINE MONLI2D(FDT,X1,X2,N1,N2,XV,FUN)}\label{p:MONLI2D}
\begin{entry}
\itemc{Purpose:} Linear 2D interpolation from the table of any function.

\iteme{DOUBLE PRECISION FDT(N1,N2)({\it input})  :}\label{p:FDTLI2D} \ttt{N1}$\times$\ttt{N2} dimensional table of values for the
input function.
\iteme{DOUBLE PRECISION X1(N1),X2(N2)({\it input})  :}\label{p:X12LI2D} arrays for discrete points corresponding to the values
of the input function.
\iteme{INTEGER N1,N2({\it input})  :}\label{p:N12LI2D} dimensions of the 2D interpolation table.
\iteme{DOUBLE PRECISION XV(2)({\it input})  :}\label{p:XVLI2D} input values for two variables of the function.
\iteme{DOUBLE PRECISION FUN({\it output})  :}\label{p:FUNLI2D} output value of the function.

\iteme{Calling by:} \ttt{MONDATA}
\end{entry}

\drawbox{SUBROUTINE MONLI4D(FDT,X1,X2,X3,X4,N1,N2,N3,N4,XV,FUN)}\label{p:MONLI4D}
\begin{entry}
\itemc{Purpose:} Linear 4D interpolation from the table of any function.

\iteme{DOUBLE PRECISION FDT(N1,N2,N3,N4)({\it input})  :}\label{p:FDTLI4D}  \ttt{N1}$\times$\ttt{N2}$\times$\ttt{N3}$\times$\ttt{N4} dimensional table of values for the input function.
\iteme{DOUBLE PRECISION X1(N1),X2(N2),X3(N3),X4(N4)({\it input})  :}\label{p:X12LI4D} arrays for discrete points corresponding to the 
values of the input function.
\iteme{INTEGER N1,N2,N3,N4({\it input})  :}\label{p:N12LI4D} dimensions of the 4D interpolation table.
\iteme{DOUBLE PRECISION XV(4)({\it input})  :}\label{p:XVLI4D} input values for four variables of the function.
\iteme{DOUBLE PRECISION FUN({\it output})  :}\label{p:FUNLI4D} output value of the function.

\iteme{Calling by:} \ttt{MONDATA}
\end{entry}

\drawbox{SUBROUTINE MONIN2D(FF,X1,X2,N1,N2,FF1,FF2,FF3)}\label{p:MONIN2D}
\begin{entry}
\itemc{Purpose:} calculations of additional integrated tables used in the generation subroutine \ttt{MONGE2D},\ttt{MONG2D4}.

\iteme{DOUBLE PRECISION FF(N1,N2)({\it input})  :}\label{p:FFIN2D} input table of 2D function.
\iteme{DOUBLE PRECISION X1(N1),X2(N2)({\it input})  :}\label{p:X12IN2D} arrays for discrete points corresponding to the 
values of the input function.
\iteme{INTEGER N1,N2({\it input})  :}\label{p:N12IN2D} dimensions of the 2D interpolation table.
\iteme{DOUBLE PRECISION FF1(N1),FF2(N1),FF3(2,N1,N2)({\it output})  :}\label{p:FF123IN2D} generated auxiliary tables.

\iteme{Calling by:} \ttt{MONDATA}, \ttt{MONIN4D}
\end{entry}

\drawbox{SUBROUTINE MONIN4D(FF,X1,X2,N1,N2,FF1,FF2,FF3)}\label{p:MONIN4D}
\begin{entry}
\itemc{Purpose:} calculations of additional integrated tables used in the generation subroutine \ttt{MONG2D4}.

\iteme{DOUBLE PRECISION FF(N1,N1,N2,N2)({\it input})  :}\label{p:FFIN4D}  input table of 4D function.

\iteme{DOUBLE PRECISION X1(N1),X2(N2)({\it input})  :}\label{p:X12IN4D} arrays for discrete points corresponding to the 
values of the input function.
\iteme{INTEGER N1,N2({\it input})  :}\label{p:N12IN4D} dimensions of the 4D interpolation table.
\iteme{DOUBLE PRECISION FF1(2,N1,N1,N2),FF2(2,N1,N1,N2),FF3(8,N1,N1,N2,N2)({\it output})  :}\label{p:FF123IN4D} generated auxiliary tables.

\iteme{Calling by:} \ttt{MONDATA}
\end{entry}

\drawbox{SUBROUTINE MONDATA}\label{p:MONDATA}
\begin{entry}
\itemc{Purpose:} to read tables for absorptive corrections
and for $pRn$ form factors from the external files
\ttt{Spi\_1}, \ttt{Sro\_1}, \ttt{Sa2\_1},
\ttt{S2pi\_1}, \ttt{S2ro\_1}, \ttt{S2a2\_1},
\ttt{FFpi\_1}, \ttt{FFro\_1}, \ttt{FFa2\_1}. 
These tables are used for calculation of the differential cross sections
for  SCE and DCE reactions at given energy (defined by parameter 
\ttt{MONPAR(3)}) by interpolation methods.  

\iteme{Status of call:} is called if \ttt{MONPAR(7)=1} or \ttt{MONPAR(8)=1}. 

\iteme{Calling by:} \ttt{MONINIT}

\iteme{Calling of:} \ttt{MONCUBI}, \ttt{MONLI2D}, \ttt{MONLI4D}, 
\ttt{MONIN2D}, \ttt{MONIN4D}

\end{entry}

\subsection{Main Functions}
\label{ss:mainfunctions}

\drawbox{DOUBLE PRECISION FUNCTION MONCSEC(KP,KR)}\label{p:MONCSEC}
\begin{entry}
\itemc{Purpose:} to give the value of the total cross section of SCE ($pp\to nX$) or 
DCE ($pp\to nXn$) reaction for the given reggeon exchange at the c.m.s. 
energy defined by parameter \ttt{MONPAR(3)} for the model defined by parameters \ttt{MONPAR(4)} and \ttt{MONPAR(5)}. 

\iteme{INTEGER KP ({\it input})  :}\label{p:KP}  single or double exchange 

\begin{subentry}
\iteme{= 1 :} for SCE cross section
\iteme{= 2 :} for DCE cross section
\end{subentry}

\iteme{INTEGER KR ({\it input })  :}\label{p:KR} type of the reggeon exchange  

\begin{subentry}
\iteme{= 1 :} for SCE define $\pi^+$ exchange, for DCE $\pi^+\pi^+$ one. 
\iteme{= 2 :} for SCE $\rho^+$ exchange, for DCE $\pi^+\rho^+$.
\iteme{= 3 :} for SCE $a_2^+$ exchange, for DCE $\pi^+a_2^+$.
\end{subentry}

\iteme{Calling by:} \ttt{MONINIT}
\end{entry}

\drawbox{DOUBLE PRECISION FUNCTION MONCSCE(NO,XI,QT)}\label{p:MONCSCE}
\begin{entry}
\itemc{Purpose:} to give the value of cross section of SCE ($pp\to nX$) 
reaction for the given reggeon exchange at given $\xi_n$ of neutron and 
$Q_t$ of reggeon at the c.m.s. energy defined by parameter \ttt{MONPAR(3)} 
for the model defined by parameters \ttt{MONPAR(4)} and \ttt{MONPAR(5)}. 

\iteme{INTEGER NO ({\it input})  :}\label{p:NOCSCE} type of the reggeon exchange  

\begin{subentry}
\iteme{= 1 :} for $\pi^+$ exchange. 
\iteme{= 2 :} for $\rho^+$ exchange.
\iteme{= 3 :} for $a_2^+$ exchange.
\end{subentry}

\iteme{DOUBLE PRECISION XI ({\it input })  :}\label{p:XICSCE} 
$\xi_n = \frac{|p_{beam}-p_{n}|}{p_{beam}}$, relative momentum loss of the neutron. 
 
\iteme{DOUBLE PRECISION QT ({\it input})  :}\label{p:QTCSCE}
$Q_t$, transverse momentum of the exchange reggeon.  

\iteme{Calling by:} \ttt{MONCDCE}, \ttt{MONDATA}
\end{entry}

\drawbox{DOUBLE PRECISION FUNCTION MONCDCE(NO,XI1,XI2,QT1,QT2)}\label{p:MONCDCE}
\begin{entry}
\itemc{Purpose:} to give the value of cross section of the 
DCE ($pp\to nXn$) reaction for the given reggeon exchange 
at given $\xi_n^{1,2}$ of neutron and  $Q^{1,2}_t$ of reggeons  at the c.m.s. 
energy defined by parameter \ttt{MONPAR(3)}
for the model defined by parameters \ttt{MONPAR(4)} and \ttt{MONPAR(5)}. 

\iteme{INTEGER NO ({\it input })  :}\label{p:NOCDCE}  
type of the reggeon exchange  

\begin{subentry}
\iteme{= 1 :} for $\pi^+\pi^+$ exchange. 
\iteme{= 2 :} for $\pi^+\rho^+$ exchange.
\iteme{= 3 :} for $\pi^+a_2^+$ exchange.
\end{subentry}

\iteme{DOUBLE PRECISION XI ({\it input })  :}\label{p:XICDCE} 
$\xi_n^{1,2} = \frac{|p_{beam}^{1,2}-p_n^{1,2}|}{p_{beam}^{1,2}}$, 
relative momentum loss of the neutrons. 

\iteme{DOUBLE PRECISION QT ({\it input })  :}\label{p:QTCDCE}  
$Q^{1,2}_t$, transverse momentum of the exchange reggeons.

\iteme{Calling by:} \ttt{MONCDCE}, \ttt{MONDATA}
\end{entry}

\drawbox{DOUBLE PRECISION FUNCTION MONCSRP(NO,NCSMOD,SVAR)}\label{p:MONCSRP}
\begin{entry}
\itemc{Purpose:} to give the value of the total reggeon-proton cross section

\iteme{INTEGER NO ({\it input })  :}\label{p:NOCSRP}  
type of the reggeon exchange  

\begin{subentry}
\iteme{= 1 :} for $\pi^+$ exchange. 
\iteme{= 2 :} for $\rho^+$ exchange.
\iteme{= 3 :} for $a_2^+$ exchange.
\end{subentry}

\iteme{INTEGER NCSMOD ({\it input })  :}\label{p:NCSMODRP} 
type of  model for the reggeon-proton cross section calculation

\begin{subentry}
\iteme{= 1 :} Donnachie-Landshoff parametrization~\cite{landshofftot}. 
\iteme{= 2 :} COMPETE parametrization~\cite{COMPETE}. 
\iteme{= 3 :} Bourrely-Soffer-Wu parametrization~\cite{BSW}.
\iteme{= 4 :} Godizov-Petrov parametrization~\cite{godizov1}.
\end{subentry}

\iteme{DOUBLE PRECISION SVAR ({\it input })  :}\label{p:SVARRP}  
invariant mass of reggeon-proton system

\iteme{Calling by:} \ttt{MONCSRR}, \ttt{MONCSCE}, \ttt{MONCDCE}
\end{entry}

\drawbox{DOUBLE PRECISION FUNCTION MONCSRR(NO,NCSMOD,SVAR)}\label{p:MONCSRR}
\begin{entry}
\itemc{Purpose:} to give the value of the total reggeon-reggeon cross section

\iteme{INTEGER NO ({\it input })  :}\label{p:NOCSRR}  
type of the reggeon-reggeon exchange  

\begin{subentry}
\iteme{= 1 :} for $\pi^+\pi^+$ exchange. 
\iteme{= 2 :} for $\pi^+\rho^+$ exchange.
\iteme{= 3 :} for $\pi^+a_2^+$ exchange.
\end{subentry}

\iteme{INTEGER NCSMOD ({\it input })  :}\label{p:NCSMODRR} 
type of  model for the reggeon-reggeon cross section calculation

\begin{subentry}
\iteme{= 1 :} Donnachie-Landshoff parametrization~\cite{landshofftot}. 
\iteme{= 2 :} COMPETE parametrization~\cite{COMPETE}. 
\iteme{= 3 :} Bourrely-Soffer-Wu parametrization~\cite{BSW}.
\iteme{= 4 :} Godizov-Petrov parametrization~\cite{godizov1}.
\end{subentry}

\iteme{DOUBLE PRECISION SVAR ({\it input })  :}\label{p:SVARRR}  
invariant mass of reggeon-reggeon system

\iteme{Calling by:} \ttt{MONDATA}, \ttt{MONCDCE}
\end{entry}

\subsection{Main Commonblocks and Parameters}
\label{ss:maincommonblocks}

\drawboxthree
{PARAMETER   (MXGLPAR=200)}
{REAL MONPAR}
{COMMON/MONGLPA/ MONPAR(MXGLPAR)}
\label{p:MONGLPA}

\begin{entry}
\itemc{Purpose:} to give access to the main \Mo switches and parameters
 
\boxsep
 
\iteme{MONPAR(1) :}\label{p:MONPAR} 
number of events for the generation.
 
\iteme{MONPAR(2) :}
switch for LHE output.
\begin{subentry}
\iteme{= 0 :} LHE output is switched off.
\iteme{= 1 :} LHE output is switched on.
\end{subentry}
 
\iteme{MONPAR(3) :}
pp centre mass energy, in GeV, (from 900 to 14000 GeV).
 
\iteme{MONPAR(4) :}
kod of model for pR and RR interaction.
\begin{subentry}
\iteme{= 1 :} Donnachie-Landshoff parametrization~\cite{landshofftot}. 
\iteme{= 2 :} COMPETE parametrization~\cite{COMPETE}. 
\iteme{= 3 :} Bourrely-Soffer-Wu parametrization~\cite{BSW}.
\iteme{= 4 :} Godizov-Petrov parametrization~\cite{godizov1}.
\end{subentry}
 
\iteme{MONPAR(5) :}
kod of model for absorptive corrections.
\begin{subentry}
\iteme{= 1 :} 3 Pomerons eikonal model~\cite{3Pomerons}. 
\end{subentry}
 
\iteme{MONPAR(6) :}
type of exchange reggeon.
\begin{subentry}
\iteme{= 1 :} for SCE define $\pi^+$ exchange, 
for DCE $\pi^+\pi^+$ one. 
\iteme{= 2 :} for SCE $\rho^+$ exchange, for DCE $\pi^+\rho^+$.
\iteme{= 3 :} for SCE $a_2^+$ exchange, for DCE $\pi^+a_2^+$.
\end{subentry}
 
\iteme{MONPAR(7) :}
switch for SCE generation.
\begin{subentry}
\iteme{= 0 :} SCE is switched off.
\iteme{= 1 :} SCE is switched on.
\end{subentry}
 
\iteme{MONPAR(8) :}
switch for DCE generation.
\begin{subentry}
\iteme{= 0 :} DCE is switched off.
\iteme{= 1 :} DCE is switched on.
\end{subentry}

\begin{subentry} 
\itemc{Note 1:}
if \ttt{MONPAR(7)=0} and \ttt{MONPAR(8)=0}, 
minimum bias events are generated by \Py.
\itemc{Note 2:} in the present version of \Mo, v.1.1, 
the simultaneous generation of SCE and DCE is impossible.
\end{subentry}

\end{entry}

\drawboxthree
{DOUBLE PRECISION S}
{INTEGER NMODPP,NMODRR,ITYPR}
{COMMON/MONTAB0/S,NMODPP,NMODRR,ITYPR}
\label{p:MONTAB0}

\begin{entry}
\itemc{Purpose:} to give access to some important \Mo\ parameters.
 
\boxsep

\iteme{S :}\label{p:STAB0}  
pp c.m.s. energy, in GeV.

\iteme{NMODPP :}\label{p:NMODPPTAB0}
kod of model for absorptive corrections.
\begin{subentry}
\iteme{= 1 :} 3 Pomerons eikonal model~\cite{3Pomerons}. 
\end{subentry}

\iteme{NMODRR :}\label{p:NMODRRTAB0}
kod of model for pR and RR interaction.
\begin{subentry}
\iteme{= 1 :} Donnachie-Landshoff parametrization~\cite{landshofftot}. 
\iteme{= 2 :} COMPETE parametrization~\cite{COMPETE}. 
\iteme{= 3 :} Bourrely-Soffer-Wu parametrization~\cite{BSW}.
\iteme{= 4 :} Godizov-Petrov parametrization~\cite{godizov1}.
\end{subentry}

\iteme{ITYPR :}\label{p:ITYPRTAB0}
type of exchange reggeon.
\begin{subentry}
\iteme{= 1 :} for SCE define $\pi^+$ exchange, 
for DCE $\pi^+\pi^+$ one. 
\iteme{= 2 :} for SCE $\rho^+$ exchange, for DCE $\pi^+\rho^+$.
\iteme{= 3 :} for SCE $a_2^+$ exchange, for DCE $\pi^+a_2^+$.
\end{subentry}

\end{entry}

\drawboxtwo
{DOUBLE PRECISION XSQ,PI,MPI,MP,MN,MRHO,MA2}
{COMMON/MONTAB1/XSQ(6),PI,MPI,MP,MN,MRHO,MA2}
\label{p:MONTAB1}

\begin{entry}
\itemc{Purpose:} to give access to some important \Mo\ parameters.
 
\boxsep

\iteme{XSQ :} six values of $\sqrt{s}$ for the interpolation subroutine \ttt{MONCUBI}.
\iteme{PI :} \ttt{3.141592653589793D0}
\iteme{MPI :} pion mass.
\iteme{MP :} proton mass.
\iteme{MN :} neutron mass.
\iteme{MRHO :} $\rho$ meson mass.
\iteme{MA2 :} $a_2$ meson mass.

\end{entry}

\drawboxtwo
{DOUBLE PRECISION XIMIN,XIMAX,QTMIN,QTMAX}
{COMMON/MONTAB2/XIMIN,XIMAX,QTMIN,QTMAX}
\label{p:MONTAB2}
\begin{entry}
\itemc{Purpose:} to give access to some important \Mo\ parameters.
 
\boxsep

\iteme{XIMIN :} minimal value of the variable $\xi$.
\iteme{XIMAX :} maximal value of the variable $\xi$.
\iteme{QTMIN :} minimal value of the variable $|\mbox{\bf q}|$ (transverse momentum of the neutron).
\iteme{QTMAX :} maximal value of the variable $|\mbox{\bf q}|$ (transverse momentum of the neutron).

\end{entry}

\drawboxtwo
{DOUBLE PRECISION API,ARHO,AA2,R2PI,R2RHO,R2A2}
{COMMON/MONTAB3/API,ARHO,AA2,R2PI,R2RHO,R2A2}
\label{p:MONTAB3}
\begin{entry}
\itemc{Purpose:} to give access to some important \Mo\ parameters.
 
\boxsep

\iteme{API :} slope of the pion regge trajectory.
\iteme{ARHO :} slope of the $\rho$ meson regge trajectory.
\iteme{AA2 :} slope of the $a_2$ meson regge trajectory.
\iteme{R2PI :} slope of the exponent in the residue of the pion trajectory.
\iteme{R2RHO :} slope of the exponent in the residue of the $\rho$ meson trajectory.
\iteme{R2A2 :} slope of the exponent in the residue of the $a_2$ meson trajectory.

\end{entry}

\drawboxtwo
{DOUBLE PRECISION GPI,GRHO,GA2,SIGRSQ,KARHO,KAA2}
{COMMON/MONTAB4/GPI,GRHO,GA2,SIGRSQ,KARHO,KAA2}
\label{p:MONTAB4}
\begin{entry}
\itemc{Purpose:} to give access to some important \Mo\ parameters.
 
\boxsep

\iteme{GPI,GRHO,GA2 :}  constants $G^2_{\pi^+pn}/(8\pi)$, $\tilde{G}^2_{\rho^+pn}/(8\pi)$ and $\tilde{G}^2_{a_2^+pn}/(8\pi)$.
\iteme{SIGRSQ :} $|\eta_R|^2$.
\iteme{KARHO, KAA2 :} $\kappa_{\rho}$, $\kappa_{a_2}$.

\end{entry}

\drawboxseven
{\ \ DOUBLE PRECISION SSPI,SSRHO,SSA2,SDPIS,SDPIA,}
{\& SDRHOS,SDRHOA,SDA2S,SDA2A,FFDPI,FFDRHO,FFDA2}     
{\ \ COMMON/MONDGET/SSPI(6,53,41),SSRHO(6,53,51),}
{\& SSA2(6,53,51),SDPIS(6,10,8,9,8),SDPIA(6,10,8,9,8),}
{\& SDRHOS(6,10,8,9,8),SDRHOA(6,10,8,9,8),SDA2S(6,10,8,9,8),}     
{\& SDA2A(6,10,8,9,8),FFDPI(6,60,16),FFDRHO(6,60,16),}
{\& FFDA2(6,60,16)}     
\label{p:MONDGET}
\begin{entry}
\itemc{Purpose:} to give access to the input tables.
 
\boxsep

\end{entry}

\drawboxseven
{\ \ DOUBLE PRECISION XSSPI,XSSRHO,XSSA2,XSDPIS,XSDPIA,}
{\& XSDRHOS,XSDRHOA,XSDA2S,XSDA2A,XFFDPI,XFFDRHO,XFFDA2}     
{\ \ COMMON/MONDFIX/XSSPI(53,41),XSSRHO(53,51),}
{\& XSSA2(53,51),XSDPIS(10,8,9,8),XSDPIA(10,8,9,8),}
{\& XSDRHOS(10,8,9,8),XSDRHOA(10,8,9,8),XSDA2S(10,8,9,8),}     
{\& XSDA2A(10,8,9,8),XFFDPI(60,16),XFFDRHO(60,16),}
{\& XFFDA2(60,16)}     
\label{p:MONDFIX}
\begin{entry}
\itemc{Purpose:} to give access to the additional tables obtained from the input files.
 
\boxsep

\end{entry}

\drawboxfour
{\ \ DOUBLE PRECISION SPI,SRHO,SA2,DPI,DRHO,DA2,FDPI,FDRHO,FDA2}
{\ \ COMMON/MONDMOD/SPI(41,41),SRHO(41,41),SA2(41,41),}
{\& DPI(17,17,17,17),DRHO(17,17,17,17),DA2(17,17,17,17),}
{\& FDPI(17,17),FDRHO(17,17),FDA2(17,17)}     
\label{p:MONDMOD}
\begin{entry}
\itemc{Purpose:} to give access to the tables for 2D and 4D generations.
 
\boxsep

\end{entry}

\drawboxnine
{\ \ DOUBLE PRECISION SPIX1,SPIX2,SPIXQ,SROX1,SROX2,SROXQ,}
{\& SA2X1,SA2X2,SA2XQ,DPIX1,DPIX2,DPIXX,}
{\& DROX1,DROX2,DROXX,DA2X1,DA2X2,DA2XX}
{\ \ COMMON/MONDGE1/SPIX1(41),SPIX2(41),SPIXQ(2,41,41),}
{\& SROX1(41),SROX2(41),SROXQ(2,41,41),}
{\& SA2X1(41),SA2X2(41),SA2XQ(2,41,41),}
{\& DPIX1(17),DPIX2(17),DPIXX(2,17,17),}
{\& DROX1(17),DROX2(17),DROXX(2,17,17),}
{\& DA2X1(17),DA2X2(17),DA2XX(2,17,17)}
\label{p:MONDGE1}
\begin{entry}
\itemc{Purpose:} to give access to the auxiliary tables for 2D and 4D generations.
 
\boxsep

\end{entry}

\drawboxsix
{\ \ DOUBLE PRECISION DDPI1,DDPI2,DDPI3,DDRO1,DDRO2,DDRO3,}
{\& DDA21,DDA22,DDA23}
{\ \ COMMON/MONDGE2/}
{\& DDPI1(4,17,17,17),DDPI2(4,17,17,17),DDPI3(8,17,17,17,17),}
{\& DDRO1(4,17,17,17),DDRO2(4,17,17,17),DDRO3(8,17,17,17,17),}
{\& DDA21(4,17,17,17),DDA22(4,17,17,17),DDA23(8,17,17,17,17)}     
\label{p:MONDGE2}
\begin{entry}
\itemc{Purpose:} to give access to the auxiliary tables for 2D and 4D generations.
 
\boxsep

\end{entry}

\drawboxfive
{\ \ DOUBLE PRECISION VXIR,VFIS,VFIA,VQTR,}
{\& VXIRF,VFIF,VXI,VQT,SVXI,SVQT,DVXI,DVQT}
{\ \ COMMON/MONDVAR/VXIR(10),VFIS(8),VFIA(8),VQTR(9),}
{\& VXIRF(60),VFIF(16),VXI(53),VQT(41),}
{\& SVXI(41),SVQT(41),DVXI(17),DVQT(17)}
\label{p:MONDVAR}
\begin{entry}
\itemc{Purpose:} to give access to the arrays of variables for the input and auxiliary tables.
 
\boxsep

\end{entry}

\newpage

\section{Program Installation}
\label{s:install}
 
Some materials related to the \Mo\ physics and generator  is the one found 
on the web page\\[2mm]
\drawbox{\ttt{http://rioutine.web.cern.ch/rioutine}}\\
in the section "Generators".
To get the code of the generator one should download the file \\[2mm]  
\drawbox{\ttt{http://rioutine.web.cern.ch/rioutine/gencode/moncher1.1.tar.gz}}\\
The program is written essentially entirely in standard Fortran 77,
and should run on any platform with such a compiler. 

The following installation procedure is suggested for the Linux users,
it was tested with CERN SLC5. 
\begin{verbatim} 
$ gunzip moncher1.1.tar.gz
$ tar -cvf  moncher1.1.tar
$ cd moncher/1.1.0
$ ls
\end{verbatim} 

\noindent
Now you can see some files:\vskip 1mm
\begin{entry}
\iteme{README} contains brief description of the files in the current directory;
\iteme{moncher.f} is the code of the generator;
\iteme{moncher.par} defines switch keys and parameters for the simulation;
\iteme{Spi\_1 Sro\_1 Sa2\_1 } 
contain data for the calculations of absorptive corrections for SCE; 
\iteme{Sp2i\_1 S2ro\_1 S2a2\_1} 
contain data for absorptive corrections for DCE;
\iteme{FFpi\_1 FFro\_1 FFa2\_1} 
contain data for form-factors;
\iteme{mkmoncher} is the executable file to compile and link \ttt{moncher.f};
\iteme{rmoncher} is the executable file to run \ttt{moncher} created by 
                 \ttt{mkmoncher}.
\end{entry}

\begin{verbatim} 
$ ./mkmoncher
\end{verbatim} 
compiles \ttt{moncher.f} by $g77$ compiler and link the generator with 
\Py\ 6.420~\cite{pythia} and some CERNLIB libraries. Then, created executable
\ttt{moncher} should be run by 
\begin{verbatim} 
$ ./rmoncher
\end{verbatim} 
Result of the simulation should be the \Py standard listing of one generated
event of the SCE reaction $pp\to nX$ at c.m.s. energy 7 TeV. The listing should 
be printed on the screen. 
If you have passed successfully all above, 
get start with the next step.

\newpage
\section{Getting Started with the Simple Example}
\label{s:start}

The Simple Example could look as following:

\begin{verbatim} 
      PROGRAM  MAIN 
      IMPLICIT DOUBLE PRECISION(A-H, O-Z)
      IMPLICIT INTEGER(I-N)
c...global MONCHER parameters
      INTEGER MXGLPAR      
      REAL MONPAR
      PARAMETER   (MXGLPAR=200)
      COMMON/MONGLPA/ MONPAR(MXGLPAR)

c...initialization      
      CALL MONGIVE('MONPAR(1)=1000')   ! number of events 
      CALL MONGIVE('MONPAR(2)=1')      ! switch for LHE saving
      CALL MONGIVE('MONPAR(3)=7000')   ! pp centre mass energy in GeV
      CALL MONGIVE('MONPAR(4)=1')      ! code of model for pR/RR interaction
      CALL MONGIVE('MONPAR(5)=1')      ! code of model for absorption
      CALL MONGIVE('MONPAR(6)=1')      ! type of Reggeon
      CALL MONGIVE('MONPAR(7)=1')      ! switch for SCE generation
      CALL MONGIVE('MONPAR(8)=0')      ! switch for DCE generation
      CALL MONGIVE('MSEL=2')           ! pythia: mb+sd+dd+elastic+lowpt
      CALL MONINIT
      
      NTOT=MONPAR(1)       
      KLHE=MONPAR(2)
      
c...generation             
      DO NEV=1,NTOT
        CALL MONEVEN
        IF(NEV.EQ.1)  CALL PYLIST(1)
        CALL ANALYZER(IOUT)
        IF(KLHE.EQ.1.AND.IOUT.EQ.1) CALL MONUPEV	
      ENDDO 

c...final statistics
      CALL PYSTAT(1)

c...produce final Les Houches Event File.
      IF(KLHE.EQ.1) CALL PYLHEF
    
      STOP
      END            
\end{verbatim} 

First, we set some values for elements of array \ttt{MONPAR} which control
a process of generation. Then, we should initialize the generator calling \ttt{MONINIT}. 
In this example we are going to generate 1000 events
of Single Pion Exchange, $pp\to n (\pi^+ p)\to n X$, at c.m.s. energy 7 TeV.
The $(\pi^+ p)$ interaction is controlled by \Py\, and it includes minimum bias,
single and double diffraction, elastic scattering and low-pt scattering.
Filling of \ttt{MONPAR} elements can be done also from the external file 
\ttt{moncher.par}. Subroutine \ttt{MONPARA} calling by \ttt{MONINIT} 
checks the presence of the \ttt{moncher.par} in the current directory
and, if it exists, reads parameters \ttt{MONPAR}, see chapter~\ref{s:control}.

On the next step, we generate  some number of events, defined by \ttt{MONPAR(2)}.
Every event is generated by \ttt{MONEVEN}.  
User's subroutine \ttt{ANALYZER(IOUT)} is called after every event generation,
analyses the event and sets some value to the integer variable \ttt{IOUT}
If \ttt{IOUT} is equal to unity, we save this event in the LHE format 
using the subroutine \ttt{MONUPEV}.

Here you can see example of the Simple Analyzer:
\begin{verbatim} 
      SUBROUTINE ANALYZER(IOUT) 
      IMPLICIT DOUBLE PRECISION(A-H, O-Z)
      IMPLICIT INTEGER(I-N)
c...HEPEVT commonblock.
      PARAMETER (NMXHEP=4000)
      COMMON/HEPEVT/NEVHEP,NHEP,ISTHEP(NMXHEP),IDHEP(NMXHEP),
     &JMOHEP(2,NMXHEP),JDAHEP(2,NMXHEP),PHEP(5,NMXHEP),VHEP(4,NMXHEP)
      DOUBLE PRECISION PHEP,VHEP
      SAVE /HEPEVT/
c
      IOUT    =0
      ISIGN   =1
      NEUTRONS=0
c      
      CALL PYHEPC(1)
c             
      DO I=1,NHEP
       KP  =IDHEP(I)
       ETA =PYP(I,19)       
       IF(KP.EQ.2112.AND.DABS(ETA).GE.8.5) THEN
         NEUTRONS=NEUTRONS+1
         ISIGN=ISIGN*ETA
       ENDIF 
      ENDDO 
c
      IF(NEUTRONS.EQ.2.AND.ISIGN.LT.0) IOUT=1
c    
      RETURN
      END            
\end{verbatim}
In this example, we analyse all particles in the generated event and look for
the neutrons (code 2112) in the region of pseudorapidity $|\eta|\ge 8.5$
(assumed acceptance of the neutron detector).
If number of such neutrons is equal to 2 and they move in opposite 
directions, \ttt{IOUT} is set to unity.     
 
Finally, we print the \Py statistics by \ttt{PYSTAT} and produce the final LHE file
which has the name \ttt{moncher.lhe} by default.

This example has a concrete physical meaning. We have selected SCE events 
with 2 leading neutrons moving in the opposite directions which imitate a DCE process.
So, we have saved background for the DCE from the SCE.      

\newpage
\section{Program Control Parameters}
\label{s:control}
All parameters that control the generation can be defined in the external file \ttt{moncher.par}.
For example, the set of parameters for the generation of the S$\pi$E process, 
described in the chapter~\ref{s:start}, can look as follows:
\begin{verbatim}
c--------------------- MONCHER v.1.1.0 card file
c
c----------------------------------------------- MONCHER control keys
c
MONPAR(1)=1000         ! number of events to generate
c
MONPAR(2)=1            ! key for Les Houches data(1-save,0-no)
c
MONPAR(3)=7000         ! pp centre mass energy in GeV  (900 -> 14000)
c
MONPAR(4)=1            ! code of model for pR and RR interaction
c                        NMODRR=1 -> Donnachie-Landshoff model (default)
c                        NMODRR=2 -> COMPETE (PDG) model
c                        NMODRR=3 -> Bourreli-Sopfer-Wu model
c                        NMODRR=4 -> Godizov-Petrov model
c
MONPAR(5)=1            ! code of model for absorption
c                        NMODPP=1  3 IP eikonal model (default)
c     not now                   NMODPP=2 -> Godizov-Petrov model 
c     not now                   NMODPP>2 -> other models...
c
MONPAR(6)=1            ! type of Reggeon (1-pi+, 2-rho+, 3-a2+) 
c                      (for DCE only pi-pi, pi-rho and pi-a2 survive)
c
MONPAR(7)=1            ! key for SCE generation
c
MONPAR(8)=0            ! key for DCE generation
c
c----------------------------------------------- PYTHIA control keys
c
cMSEL =0                 ! full user control
cMSUB(11)=1              ! f + f' -> f + f' (QCD)
cMSUB(12)=1              ! f + fbar -> f' + fbar'
cMSUB(13)=1              ! f + fbar -> g + g
cMSUB(28)=1              ! f + g -> f + g 
cMSUB(53)=1              ! g + g -> f + fbar
cMSUB(68)=1              ! g + g -> g + g
cMSUB(91)=1              ! Elastic scattering
cMSUB(92)=1              ! Single diffractive (AX)
cMSUB(93)=1              ! Single diffractive (XB)
cMSUB(94)=1              ! Double  diffractive
cMSUB(95)=1              ! Low-pT scattering
cMSEL =1                 ! mb
MSEL =2                 ! mb+sd+dd+elastic+lowpt
c
MRPY(1)=12031967        ! start point of random number generator
\end{verbatim} 

Subroutine \ttt{MONPARA} reads lines from \ttt{moncher.par}.
All lines begining with a letter "c" are ignored by the program, all others lines are processed 
by subroutine \ttt{MONGIVE}, which can recognize any variables from the \Mo common block \ttt{/MONGPGL/}
and the \Py common blocks 
\ttt{/PYJETS/, /PYDAT1/, /PYDAT2/, /PYDAT3/, /PYDAT4/, /PYDATR/, /PYSUBS/, /PYPARS/, /PYINT1/,
     /PYINT2/, /PYINT3/, /PYINT4/, /PYINT5/, /PYINT6/, /PYINT7/, /PYINT8/, /PYMSSM/, /PYMSRV/,
     /PYTCSM/, /PYPUED/}, (see \cite{pythia}).
Parameters \ttt{MONPARA} are described in detail in the section~\ref{s:programoverview}, 
page~\pageref{p:MONPAR}. 

Using parameters from the common blocks listed above, one can define wide spectrum of SCE  
(\ttt{MONPAR(7)=1}) and DCE ((\ttt{MONPAR(8)=1}) processes or any processes existing in \Py\  
(if (\ttt{MONPAR(7)=0} and \ttt{MONPAR(8)=0}). Some examples are described in the next 
chapter.        

\section{Examples of the \Mo\ Processes.}
\label{s:examples}
\begin{center}
\begin{table}[h!]
\begin{center}
\begin{tabular}{|c|c|c|c|c|}
\hline
N 
& Process 
& \begin{minipage}[h]{0.2\textwidth}\begin{center} Type of $\pi^+p$ \\ interactions\end{center}\end{minipage}
& Picture of the process
& \begin{minipage}[h]{0.16\textwidth}\begin{center}\vspace{1mm} The \Mo\\ parameters\vspace{1mm} \end{center}\end{minipage}\\
\hline
1
& $pp\to nX$ 
& \begin{minipage}[h]{0.2\textwidth}\begin{center} 
minimum bias: \\ $\pi^+p\to X$ 
\end{center}\end{minipage} 
& \begin{minipage}[h]{0.25\textwidth}\centerline{\includegraphics[width=0.8\textwidth]{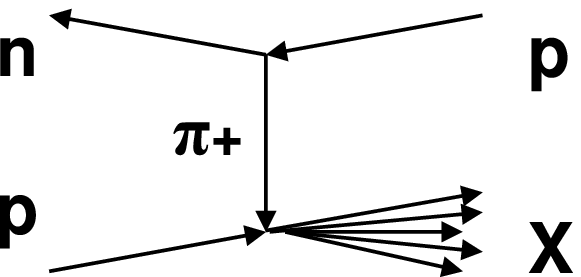}}\end{minipage} 
& \begin{minipage}[h]{0.14\textwidth}\vspace{0.3\baselineskip}
\ttt{MONPAR(7)=1}\\ \ttt{MONPAR(8)=0}\\ \ttt{MSEL=1}\\
\vspace{0.3\baselineskip}\end{minipage} \\
\hline
2
& $pp\to n\pi^+p$ 
& \begin{minipage}[h]{0.2\textwidth}\begin{center} 
elastic scattering:\\ $\pi^+p\to \pi^+p$ 
\end{center}\end{minipage} 
& \begin{minipage}[h]{0.25\textwidth}\centerline{\includegraphics[width=0.8\textwidth]{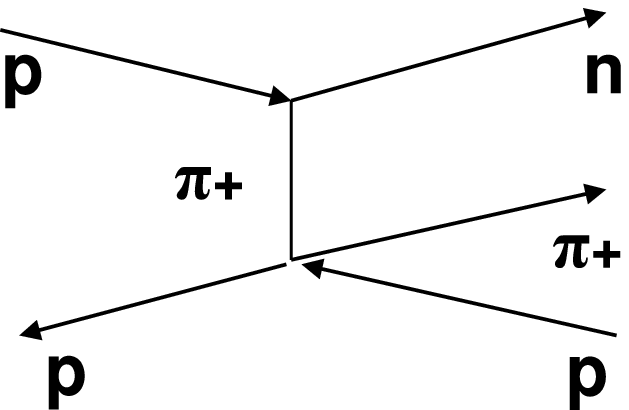}}\end{minipage} 
& \begin{minipage}[h]{0.14\textwidth}\vspace{0.3\baselineskip}
\ttt{MONPAR(7)=1}\\ \ttt{MONPAR(8)=0}\\ \ttt{MSEL=0}\\ \ttt{MSUB(91)=1}
\vspace{0.3\baselineskip}\end{minipage}\\
\hline
3
& $pp\to nXY$  
& \begin{minipage}[h]{0.2\textwidth}\begin{center} 
double diffraction:\\ $\pi^+p\to X+Y$  
\end{center}\end{minipage} 
& \begin{minipage}[h]{0.25\textwidth}\centerline{\includegraphics[width=0.8\textwidth]{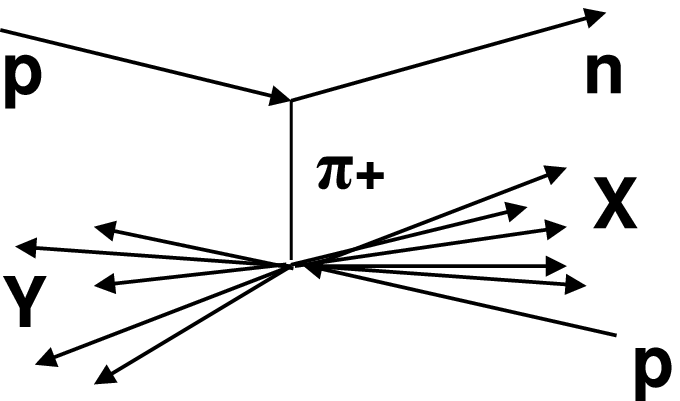}}\end{minipage} 
& \begin{minipage}[h]{0.14\textwidth}\vspace{0.3\baselineskip}
\ttt{MONPAR(7)=1}\\ \ttt{MONPAR(8)=0}\\ \ttt{MSEL=0}\\ \ttt{MSUB(94)=1}
\vspace{0.3\baselineskip}\end{minipage}\\
\hline
4
& $pp\to nXp$  
& \begin{minipage}[h]{0.2\textwidth}\begin{center} 
single diffraction\\ ($\pi^+$ dissociation):\\ $\pi^+p\to X+p$  
\end{center}\end{minipage} 
& \begin{minipage}[h]{0.25\textwidth}\centerline{\includegraphics[width=0.8\textwidth]{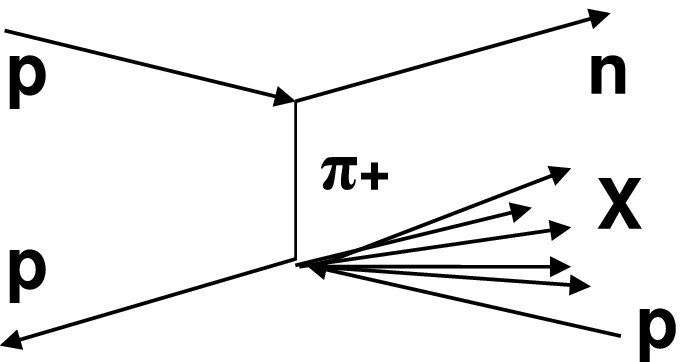}}\end{minipage} 
& \begin{minipage}[h]{0.14\textwidth}\vspace{0.3\baselineskip}
\ttt{MONPAR(7)=1}\\ \ttt{MONPAR(8)=0}\\ \ttt{MSEL=0}\\ \ttt{MSUB(92)=1}
\vspace{0.3\baselineskip}\end{minipage}\\
\hline
5
& $pp\to nX\pi^+$  
& \begin{minipage}[h]{0.2\textwidth}\begin{center} 
single diffraction\\ ($p$ dissociation):\\ $\pi^+p\to X+\pi^+$  
\end{center}\end{minipage} 
& \begin{minipage}[h]{0.25\textwidth}\centerline{\includegraphics[width=0.8\textwidth]{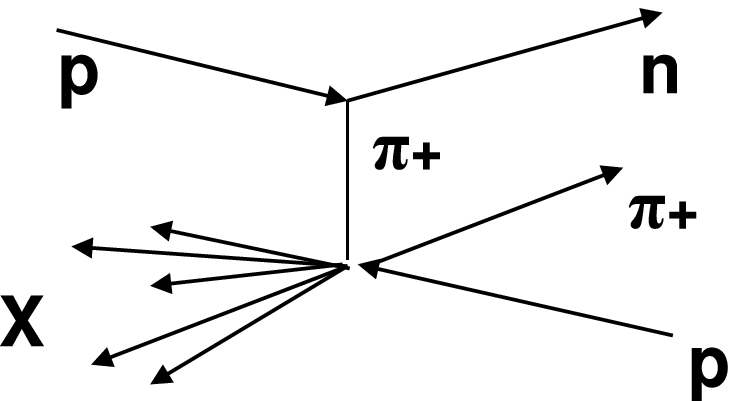}}\end{minipage} 
& \begin{minipage}[h]{0.14\textwidth}\vspace{0.3\baselineskip}
\ttt{MONPAR(7)=1}\\ \ttt{MONPAR(8)=0}\\ \ttt{MSEL=0}\\ \ttt{MSUB(93)=1}
\vspace{0.3\baselineskip}\end{minipage}\\
\hline
\end{tabular}
\caption{\label{tab:SCEpro}Some S$\pi$E processes which can be generated with \Mo.}
\end{center}
\end{table}  
\end{center}

It was mentioned already in Chapter~\ref{s:programoverview} that the \Mo\ generates $p\reg n$ 
vertices and, then, $\reg p$ (for S$\reg$E) or $\reg\reg$ (for D$\reg$E) interactions are generated by \Py. 
The type of these interactions can be controled by the \Py parameters. We can define elastic 
or inelastic interactions, diffractive or non-diffractive processes, different types of diffraction, 
hard scattering, etc. Some of the basic processes for S$\pi$E and D$\pi$E, 
which can be generated by the \Mo, are presented in the tables~\ref{tab:SCEpro} and \ref{tab:DCEpro}
respectively. 

Let us consider one more simple example, how to generate process number 2 from 
Table~\ref{tab:SCEpro}. This is a Single Pion Exchange with elastic scattering of 
the virtual pion by the proton of the beam. This reaction, $pp\to n\pi^+p$, 
has very clear signature: neutron, proton, single $\pi^+$ meson and nothing else 
in the final state. Initial particles are scattered at very small angles and, thereof, there are
no any detector signals in the region of pseudorapidity $|\eta|<7$. An experimental 
possibility of such measurements has been analysed in Ref.~\cite{ourneutronel} with 
prereleased version of \Mo. 

File \ttt{moncher.par} with parameters for the generation of $pp\to n\pi^+p$ can look as follows:
\begin{verbatim}
MONPAR(1)=1            ! number of events to generate
MONPAR(2)=0            ! key for Les Houches data(1-save,0-no)
MONPAR(3)=7000         ! pp centre mass energy in GeV  (900 -> 14000)
MONPAR(4)=1            ! code of model for pR and RR interaction
MONPAR(5)=1            ! code of model for absorption
MONPAR(6)=1            ! type of Reggeon (1-pi+, 2-rho+, 3-a2+) 
MONPAR(7)=1            ! key for SCE generation
MONPAR(8)=0            ! key for DCE generation
MSEL =0                ! full user control
MSUB(91)=1             ! elastic scattering
\end{verbatim}   

\noindent
Parameter \ttt{MONPAR(7)=1} defines the generation of the S$\reg$E process. 
Exchange reggeon is a pion (\ttt{MONPAR(6)=1}). \Py parameters  \ttt{MSEL=0} and 
\ttt{MSUB(91)=1} set elastic $\pi^+ p$ scattering. Parameter \ttt{MONPAR(4)=1}  sets 
Donnachie-Landshoff parametrization for $\pi^+ p$ interaction, see 
subsection~\ref{sss:parametrization}. Parameter \ttt{MONPAR(5)=1} specifies 3 Pomeron 
model for absorptive correcttions, see subsection~\ref{sss:absorption}.
Parameters \ttt{MONPAR(1)=1} and \ttt{MONPAR(3)=7000} set the generation of 1 event at 7 TeV 
pp c.m.s. energy. We don't ask to save any events (\ttt{MONPAR(2)=0}) and the only 
result of the generation is the \Py listing of the generated event:
\begin{verbatim}
                            Event listing (summary)

I particle/jet KS     KF  orig    p_x      p_y      p_z       E        m

1 !p+!         21    2212    0    0.000    0.000 3500.000 3500.000    0.938
2 !p+!         21    2212    0    0.000    0.000-3500.000 3500.000    0.938
===========================================================================
3 n0            1    2112    2    0.114    0.216-2296.804 2296.804    0.940
4 !pi+!        21     211    2   -0.114   -0.216-1203.196 1203.196    0.140
===========================================================================
5 !p+!         21    2212    3   -0.019   -0.001 3500.000 3500.000    0.938
6 !pi+!        21     211    4   -0.095   -0.215-1203.196 1203.196    0.140
===========================================================================
7 p+            1    2212    5   -0.019   -0.001 3500.000 3500.000    0.938
8 pi+           1     211    6   -0.095   -0.215-1203.196 1203.196    0.140
               sum:  2.00         0.000    0.000    0.000 7000.000 7000.000
	       
\end{verbatim}   

\noindent
In this listing lines 1 and 2 correspond to the protons of the beams.
Lines 3, 7 and 8 relate to the neutron, proton and pion, respectively, in the final
state of the reaction. The proton is deflected at angle $\approx$5.5x10$^{-6}$ rad.,
neutron and pion are scattered  in the direction opposite to proton, as it is shown on the
diagram of the process in the table~\ref{tab:SCEpro}, with polar angles 
$\approx$10$^{-4}$ and $\approx$2x10$^{-4}$ rad.

\begin{center}
\begin{table}[t!]
\begin{center}
\begin{tabular}{|c|c|c|c|c|}
\hline
N 
& Process 
& \begin{minipage}[h]{0.2\textwidth}\begin{center} Type of $\pi^+\pi^+$ \\ interactions\end{center}\end{minipage}
& Picture of the process 
& \begin{minipage}[h]{0.16\textwidth}\begin{center}\vspace{1mm} The \Mo\\ parameters\vspace{1mm} \end{center}\end{minipage}\\
\hline
1
& $pp\to nXn$ 
& \begin{minipage}[h]{0.2\textwidth}\begin{center} 
minimum bias: \\ $\pi^+\pi^+\to X$ 
\end{center}\end{minipage} 
& \begin{minipage}[h]{0.25\textwidth}\centerline{\includegraphics[width=0.7\textwidth]{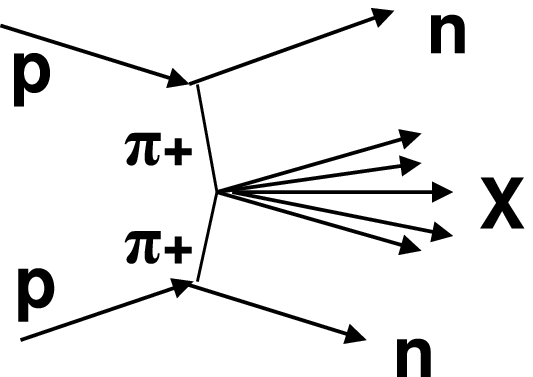}}\end{minipage} 
& \begin{minipage}[h]{0.14\textwidth}\vspace{\baselineskip}
\ttt{MONPAR(7)=0}\\ \ttt{MONPAR(8)=1}\\ \ttt{MSEL=1}\\ 
\vspace{0.3\baselineskip}\end{minipage} \\
\hline
2
& $pp\to n \pi^+\pi^+ n$ 
& \begin{minipage}[h]{0.2\textwidth}\begin{center} 
elastic scattering:\\ $\pi^+\pi^+\to \pi^+\pi^+$ 
\end{center}\end{minipage} 
& \begin{minipage}[h]{0.25\textwidth}\centerline{\includegraphics[width=0.8\textwidth]{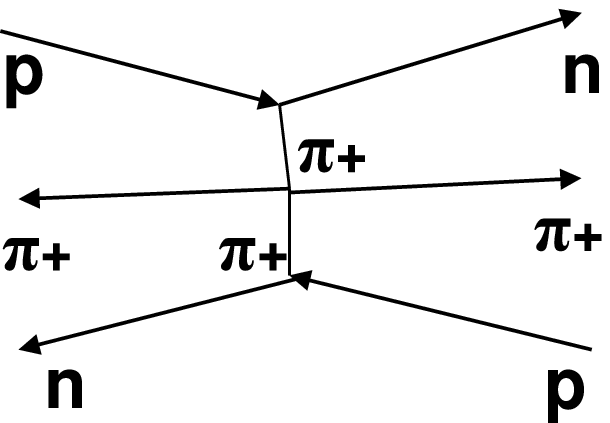}}\end{minipage} 
& \begin{minipage}[h]{0.14\textwidth}\vspace{0.3\baselineskip}
\ttt{MONPAR(7)=0}\\ \ttt{MONPAR(8)=1}\\ \ttt{MSEL=0}\\ \ttt{MSUB(91)=1}
\vspace{0.3\baselineskip}\end{minipage}\\
\hline
3
& $pp\to nXYn$  
& \begin{minipage}[h]{0.2\textwidth}\begin{center} 
double diffraction:\\ $\pi^+\pi^+\to X+Y$  
\end{center}\end{minipage} 
& \begin{minipage}[h]{0.25\textwidth}\centerline{\includegraphics[width=0.8\textwidth]{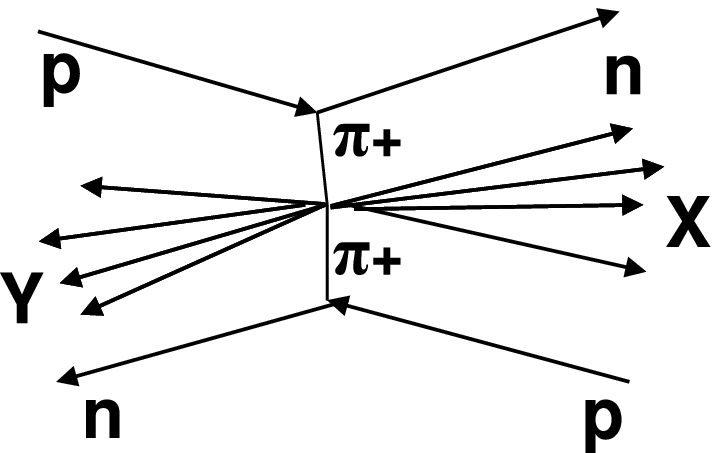}}\end{minipage} 
& \begin{minipage}[h]{0.14\textwidth}\vspace{0.3\baselineskip}
\ttt{MONPAR(7)=0}\\ \ttt{MONPAR(8)=1}\\ \ttt{MSEL=0}\\ \ttt{MSUB(94)=1}
\vspace{0.3\baselineskip}\end{minipage}\\
\hline
4
& $pp\to nX\pi^+n$  
& \begin{minipage}[h]{0.2\textwidth}\begin{center} 
single diffraction:\\ $\pi^+\pi^+\to X+\pi^+$  
\end{center}\end{minipage} 
& \begin{minipage}[h]{0.25\textwidth}\centerline{\includegraphics[width=0.8\textwidth]{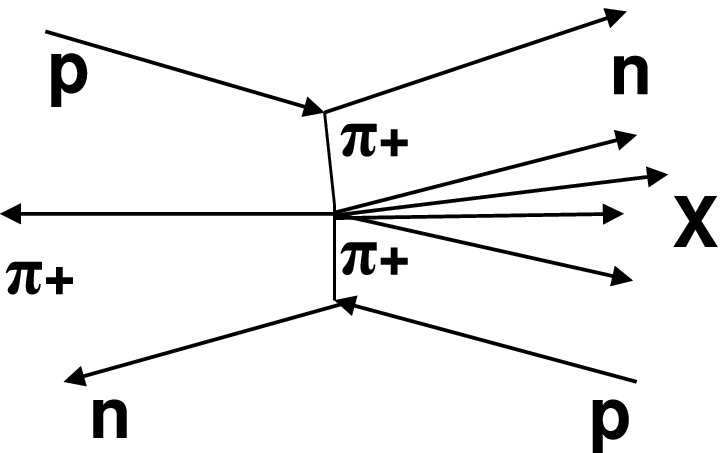}}\end{minipage} 
& \begin{minipage}[h]{0.14\textwidth}\vspace{0.3\baselineskip}
\ttt{MONPAR(7)=0}\\ \ttt{MONPAR(8)=1}\\ \ttt{MSEL=0}\\ \ttt{MSUB(92)=1}\\ or\\ \ttt{MSUB(93)=1}
\vspace{0.3\baselineskip}\end{minipage}\\
\hline
\end{tabular}
\caption{\label{tab:DCEpro}Some D$\pi$E processes which can be generated with \Mo.}
\end{center}
\end{table}  
\end{center}

\newpage
\section*{Aknowledgements}

 This work is supported by the grant RFBR-10-02-00372-a.
 

\newpage
\section*{Index of Subprograms and Common Block Variables}
\addcontentsline{toc}{section}
{Index of Subprograms and Common Block Variables}

\boxsep

\noindent
\begin{minipage}[t]{\halfpagewid}
\begin{tabular*}{\halfpagewid}[t]{@{}l@{\extracolsep{\fill}}r@{}}
\ttt{MONINIT} subroutine     & \pageref{p:MONINIT} \\
\ttt{MONTITL} subroutine     & \pageref{p:MONTITL} \\
\ttt{MONPARA} subroutine     & \pageref{p:MONPARA} \\
\ttt{MONMBDF} subroutine     & \pageref{p:MONMBDF} \\
\ttt{MONEVEN} subroutine     & \pageref{p:MONEVEN} \\
\ttt{MONSPEG} subroutine     & \pageref{p:MONSPEG} \\
\ttt{MONDPEG} subroutine     & \pageref{p:MONDPEG} \\
\ttt{MONSPEM} subroutine     & \pageref{p:MONSPEM} \\
\ttt{MONDPEM} subroutine     & \pageref{p:MONDPEM} \\
\ttt{MONSHPY} subroutine     & \pageref{p:MONSHPY} \\
\ttt{MONGIVE} subroutine     & \pageref{p:MONGIVE} \\
\ttt{MONUPEV} subroutine     & \pageref{p:MONUPEV} \\
\ttt{MONUPIN} subroutine     & \pageref{p:MONUPIN} \\
\ttt{MONGE2D} subroutine     & \pageref{p:MONGE2D} \\
\ttt{MONG2D4} subroutine     & \pageref{p:MONG2D4} \\
\ttt{MONCUBI} subroutine     & \pageref{p:MONCUBI} \\
\ttt{MONLI2D} subroutine     & \pageref{p:MONLI2D} \\
\ttt{MONLI4D} subroutine     & \pageref{p:MONLI4D} \\
\ttt{MONIN2D} subroutine     & \pageref{p:MONIN2D} \\
\ttt{MONIN4D} subroutine     & \pageref{p:MONIN4D} \\
\ttt{MONDATA} subroutine     & \pageref{p:MONDATA} \\
\ttt{MONCSEC} function       & \pageref{p:MONCSEC} \\
\ttt{MONCSCE} function       & \pageref{p:MONCSCE} \\
\ttt{MONCDCE} function       & \pageref{p:MONCDCE} \\
\ttt{MONCSRP} function       & \pageref{p:MONCSRP} \\
\ttt{MONCSRR} function       & \pageref{p:MONCSRR} \\
\ttt{MONGLPA} common block   & \pageref{p:MONGLPA} \\
\ttt{MONTAB0} common block   & \pageref{p:MONTAB0} \\
\ttt{MONTAB1} common block   & \pageref{p:MONTAB1} \\
\ttt{MONTAB2} common block   & \pageref{p:MONTAB2} \\
\ttt{MONTAB3} common block   & \pageref{p:MONTAB3} \\
\ttt{MONTAB4} common block   & \pageref{p:MONTAB4} \\
\ttt{MONDGET} common block   & \pageref{p:MONDGET} \\
\ttt{MONDFIX} common block   & \pageref{p:MONDFIX} \\
\ttt{MONDMOD} common block   & \pageref{p:MONDMOD} \\
\ttt{MONDGE1} common block   & \pageref{p:MONDGE1} \\
\ttt{MONDGE2} common block   & \pageref{p:MONDGE2} \\
\ttt{MONDVAR} common block   & \pageref{p:MONDVAR} \\
\ttt{MONPAR}      in \ttt{/MONGLPA/}       & \pageref{p:MONPARA} \\
\ttt{S}      in \ttt{/MONTAB0/}       & \pageref{p:STAB0} \\
\ttt{NMODPP} in \ttt{/MONTAB0/}       & \pageref{p:NMODPPTAB0} \\
\ttt{NMODRR} in \ttt{/MONTAB0/}       & \pageref{p:NMODRRTAB0} \\
\ttt{ITYPR}  in \ttt{/MONTAB0/}       & \pageref{p:ITYPRTAB0} \\
\end{tabular*}
\end{minipage}%

\end{document}